\documentclass[sn-mathphys]{sn-jnl}

\jyear{2021}%

\theoremstyle{thmstyleone}%
\theoremstyle{thmstyletwo}%

\theoremstyle{thmstylethree}%

\usepackage{booktabs}
\usepackage[flushleft]{threeparttable}

\graphicspath{ {figures/} }
\usepackage{dsfont} 
\usepackage{amsmath}
\newtheorem*{mydef}{Definition}
\usepackage[T1]{fontenc}

\usepackage[english]{babel}
\usepackage{algorithm}
\usepackage{algorithmicx}
\usepackage[inline]{enumitem}

\usepackage{caption}
\usepackage{subcaption}
\usepackage{multirow}
\usepackage{comment}

\newcommand{\algname}{FaiREO}
\newcommand{\vt}{\mathcal{V}_{t}}
\newcommand{\vT}{\mathcal{V}_{T}}

\newcommand{\norm}[1]{\left\lVert#1\right\rVert}

\begin{document}

\title[\algname{}: User Group Fairness for Equality of Opportunity in CR]{\algname{}: User Group Fairness for Equality of Opportunity in Course Recommendation}


\author*[1,2]{\fnm{Agoritsa} \sur{Polyzou}}\email{apolyzou@fiu.edu, \url{https://orcid.org/0000-0001-8630-7131}}

\author[1]{\fnm{Maria} \sur{Kalantzi}}\email{kalan028@umn.edu}

\author[1]{\fnm{George} \sur{Karypis}}\email{karypis@umn.edu}

\affil[1]{\orgdiv{Department of Computer Science and Engineering}, \orgname{University of Minnesota}, \orgaddress{\city{Minneapolis}, \state{MN}, \country{USA}}}

\affil[2]{\orgdiv{School of Computing and Information Sciences}, \orgname{Florida International University}, \orgaddress{\city{Miami}, \state{FL}, \country{USA}}}


\abstract{    Course selection is challenging for students in higher educational institutions. Existing course recommendation systems make relevant suggestions to the students and help them in exploring the available courses. The recommended courses can influence students' choice of degree program, future employment, and even their socioeconomic status. This paper focuses on identifying and alleviating biases that might be present in a course recommender system. We strive to promote balanced opportunities with our suggestions to all groups of students. At the same time, we need to make recommendations of good quality to all protected groups. We formulate our approach as a multi-objective optimization problem and study the trade-offs between equal opportunity and quality. We evaluate our methods using both real-world and synthetic datasets. The results indicate that we can considerably improve fairness regarding equality of opportunity, but we will introduce some quality loss. Out of the four methods we tested, GHC-Inc and GHC-Tabu are the best performing ones with different advantageous characteristics.}




\maketitle

\section{Introduction}

Higher education is valuable but it involves a significant financial cost that increases every year~\cite{snyder2019}. It is important for institutions to provide value to their students, so they often employ machine learning supporting tools. Many schools provide course recommendation systems (CRS) to facilitate course selection~\cite{basavaraj2018,jiang2019,polyzou2019,pardos2019a,esteban2018,parameswaran2011,sobecki2010}.
Existing CRSs empower learners to explore the curriculum, make informed decisions and plans while scaling advice to large cohorts~\cite{iatrellis2017,feghali2011,taha2012}. They help students to choose relevant elective courses in their curriculum according to different criteria, e.g., their individual performance, preferences, interests, and needs.
Such systems influence students' choices, degree plan, and ultimately, their career paths.


Machine learning models are built based on data, but often, this data is highly biased, resulting in outcomes that replicate existing biases~\cite{barocas2016}. Such biases can be harmful to a model, leading to discrimination against certain groups of users~\cite{mehrabi2019}. Users form different protected groups based on their protected attributes, which include gender, age, race, color, or disability. Educational data is not an exception; the simplest example is the gender bias present in historical data related to student enrollment and performance. Women have been historically underrepresented in science, technology, engineering, and mathematics, while education, health, and welfare are their most common fields of study~\cite{oecd2020}. Higher education could play a key role in improving gender equality. As course recommendation approaches get embedded in operational systems that drive decision-making, it is important to ensure that they do not discriminate against any group of users. CRSs need to be useful and beneficial to all students regardless of their protected attributes.

A body of work in recommender systems considers the case of fairness under the spectrum of multiple stakeholders~\cite{burke2016towards, mehrotra2018towards, malthouse2019multistakeholder}.  They study the benefit trade-off among the system, the vendors (i.e., the items), and the users. Other researchers consider fairness on the item side. They are interested in the diversity of items in the recommendation list of each user and attempt to impose equal exposure of different groups of items~\cite{zehlike2017fa, beutel2019fairness, biega2018equity, singh2018fairness, yang2017measuring}. Such interpretations of fairness are not sufficient to ensure the equal treatment of the students in a CRS as they do not consider the existence of different groups of users. 
Our motivation is driven by the \textit{equality of educational opportunity}~\cite{shields2017}. According to this ideal, every student should have equal educational opportunities irrespective of race, gender, socioeconomic class, sexuality, or religion. To this end, we propose \algname, a new type of fairness for course recommendation systems.
We assume that course recommendation involves a notion of \emph{opportunity}. That is, by recommending a course, a recommender system provides to a student the \emph{opportunity} to review the course's contents and consider taking it, something that they might not have done otherwise. This course could open a new path for them to explore and lead them to a job with better benefits. \algname{} promotes that each student group receives equally high-quality recommendations and equal opportunities to consider a particular course. It also alleviates the feedback loop bias occurring when users consume biased recommendations and generate biased data that are later used to generate new (biased) recommendations.

%
\algname{} operationalizes this by recommending each course at fair rates across the different student groups. We introduce four greedy hill-climbing algorithms, GHC(Gc), GHC(NoNe), GHC-Inc, and GHC-Tabu, that work in two phases. First, they make an initial assignment of recommended courses to users, and then, they refine this initial solution in order to improve the overall fairness, according to a multi-objective function. This function captures and balances two fairness-related, but often conflicting goals: equality in recommendation quality and equality of opportunity offered to students across different protected groups.



This paper's contributions include: 

(1) a new definition of fairness in recommender systems, \algname{}, that captures the equality of opportunity, 

(2) a multi-objective optimization problem formulation to consider \algname{} in course recommendation (CR), 

(3) a set of steepest-ascent hill climbing algorithms to solve this problem, 

(4) a methodology for generating synthetic datasets suitable for this problem.\\
The experimental evaluation with synthetic and real data from the University of Minnesota shows the behavior and effectiveness of the proposed algorithms. We can mitigate or even eliminate unfair recommendations w.r.t. opportunity. In the process, we may introduce unfairness in terms of unbalance in the quality of recommendations across the student groups.

The rest of the paper is organized as follows: Sect.~\ref{sec:problstatement} presents our definition of fairness and our problem statement. Sect.~\ref{sec:relwork} reviews the existing work in fairness on recommender systems, as well as other problems that could be related to our problem. In Sect.~\ref{sec:devalg}, we formulate our objective functions and present our developed algorithms. Sect.~\ref{sec:setup} details all the information regarding our experimental setup, and Sect.~\ref{sec:results} presents and analyzes our evaluation results. Finally, Sect.~\ref{sec:summary} summarizes our findings and concludes our paper.


\section{Problem Formulation}
\label{sec:problstatement}
\subsection{Assumptions and Notation}
\label{sec:assumpt}
    In this work, we make the following assumptions:
    
    \begin{itemize}
        \item [] \textbf{(Assumption 1)} The student body has at least one protected attribute based on which we can form protected groups of students (student groups). When there are more than two protected attributes that we need to consider, we create a protected group for each combination of values that they take.
    	\item [] \textbf{(Assumption 2)} We have access to a method that computes the recommendation scores for the available courses a student might take next semester. The scores accurately capture how well a course matches the student's academic level, background, and knowledge. Courses that the student has already taken receive zero recommendation scores. We will refer to the recommendation solution that suggests the $k$ highest scored courses for each student as the \textbf{HSC solution}. 
    	\item [] \textbf{(Assumption 3)} Course recommendations are fair when they are distributed proportionally to the protected groups according to their population. Alternatively, the system administrators may have insight into the desired distribution of recommendations that we consider to be fair. In any case, the fair recommendation distribution we need to achieve for each course is described by the fair distribution matrix, $\mathbf{X}$.
    \end{itemize}

    \textbf{Notation.}
    For the rest of the paper, we will adopt the following notation.
	Capital calligraphic letters will be used for sets. Lower bold case letters will indicate vectors, e.g., $\mathbf{o}$, and their elements will be denoted by regular lower case letters, e.g., $o_p$. Capital bold letters correspond to matrices, e.g., $\mathbf{Y}$, and their indexed elements will be denoted by regular lower case letters, e.g., $y_{i,j}$. We use a superscript in parenthesis to refer to the students to whom we recommended the course with the corresponding index. For example, $n^{(j)}$ is the number of students that were recommended course $j$. Table~\ref{table:notation} defines and presents the symbols we use. 

	\begin{table}[tb] 
		\centering
		\caption{Notation.}
		\begin{tabular}{ p{0.1\linewidth}| p{0.8\linewidth}} \toprule
			$i$, $j$ & Index for students, courses.\\
			$p$, $q$ & Index for student groups, course buckets.\\
			$\mathcal{S}, \mathcal{C}$ & Set of all students, courses. \\
			$n, m$ & Number of all students, courses.\\
			$g_s, g_c$ & Number of student, course subsets formed.\\
			$\mathcal{C}_{1, \dots , g_c}$ & Subset of courses. \\
			$\mathcal{S}_{1, \dots , g_s}$ & Protected groups of students. \\
			$\mathbf{Y}$ & The $(n \times m)$ recommendation score matrix of the optimal fairness-unaware model.\\
			$y_{i,j}$ & Score of course $j$ and student $i$.\\
			$\mathcal{R}_i$ & Set of recommended courses for student $i$.\\
			$k$ & Number of courses we recommend, i.e., $\lvert\mathcal{R}_i\rvert$.\\
			$\mathbf{R}$ & Recommendation solution
			$\mathbf{R}=[\mathcal{R}_1, \dots , \mathcal{R}_n]$.\\
			$n^{(j)}$ & Number of students to whom we recommend course $j$, i.e., $\lvert\{i \text{ s.t. } j \in \mathcal{R}_i\}\rvert$.\\
			$n_{p}$ & Number of students in the $\mathcal{S}_p$ group. \\
			$n^{(j)}_{p}$ & Number of students in $\mathcal{S}_p$ to whom we recommend course $j$, i.e., $\lvert\{i \in \mathcal{S}_p \text{ s.t. } j \in \mathcal{R}_i\}\rvert$.\\	
			$\mathbf{X}$ & The $(m \times g_s)$ matrix with the fair distribution of the course recommendations. \\
			$x_{j,p}$ & Fair ratio of the recommendations of course $j$ to the student group $\mathcal{S}_p$. \\
			\bottomrule
		\end{tabular}
		\label{table:notation}
	\end{table}

\subsection{Fairness in Recommendation with Equality of Opportunity}
\label{sec:faireo}
    
Course selection is often affected by existing biases and stereotypes, as well as other people's actions and opinions. As a result, course enrollment data, which is the input data of a CRS, exhibits historical, stereotype, and social biases~\cite{mehrabi2019}. 
A CRS may propagate these biases to the recommendation output. For example, such a system would rarely recommend coding classes to female students in a computer science department as computer programming is stereotypically considered a male-dominated area. 
Such a programming class could have provided Anna with the experience needed for a software engineering job. A fair CRS would ensure that students in all protected groups are offered the same opportunities; a course's recommendations are distributed proportionally to the protected groups. 

Our interest is on equal opportunity, but we still need to consider our initial goal in a CRS: support students by offering them recommendations of high quality. While these two aspects add to the value of a recommendation system, they can be in conflict. We assume that the HSC solution offers the highest quality output, but there are no guarantees for the equality of the opportunities it offers. On the other hand, if we modify the recommendation lists to satisfy equality of opportunity, some recommendations will be of lower quality. As a result, there is a need to balance these two goals. We need to ensure that the equality of opportunity does not come at the expense of the equality in offering good recommendations to the protected groups. 
    
Motivated by the above discussion, we introduce a new type of group fairness, referred to as \emph{fairness of equality of opportunity} (\algname), which is defined as follows.


\begin{mydef}
Let $\mathcal{S}$ be a population of students, that can be divided based on the value of one or more protected features into $g_s$ groups, $\mathcal{S}_{1}, \dots, \mathcal{S}_{g_s}$, with cardinalities $n_{1}, \dots, n_{g_s}$, respectively. A course recommendation system satisfies \textbf{fairness for equality of opportunity, \algname,} when:

\begin{enumerate}
    \item Each student group $p$ gets a share of each course's recommendations relative to its corresponding fair ratio, $x_{j,p}$. 
    Let $n^{(j)}_{p}$ be the number of students in group $\mathcal{S}_{p}$ to whom we recommend course $j$. Recommendations w.r.t. course $j$ offer equal opportunities when:
	\begin{equation}
		\frac{n^{(j)}_{p}}{n^{(j)}} \approx x_{j,p}, \quad\forall p \in \{1, \dots, g_s\}.
		\label{eq:deff}
	\end{equation}
    
    \item All student groups equally receive high quality recommendations with respect to the courses' recommendation scores, i.e.,
	\begin{equation}
		\sum_{i \in \mathcal{S}_p} \sum_{j \in \mathcal{R}'_i} y_{i,j} \approx \sum_{i \in \mathcal{S}_p} \sum_{j \in \mathcal{R}_i} y_{i,j}
		, \quad \forall p \in \{1, \dots, g_s\},
		\label{eq:defq}
	\end{equation}
	where $\mathcal{R}'_i$ is a set of recommended courses for student $i$, $\mathcal{R}_i$ the set of courses recommended based on the HSC solution, $y_{i,j}$ denotes the score of student $i$ in course $j$, and $\mathcal{S}_p$ denotes the students in the protected group $p$.
\end{enumerate}
\end{mydef}

Eq.~\ref{eq:deff} ensures that the distribution of a course's recommendations to the student groups matches the determined fair distribution of the recommendations to the student groups.
In Eq.~\ref{eq:defq}, we ask that the quality of a solution $\mathcal{R}'$ (measured by the sum of the recommendation scores of the courses recommended) is similar to the quality of the HSC solution for each student group $\mathcal{S}_p$. We assume that there is no underlying reason why no courses would be a good match for students in a specific student group. As a result, the quality of HSC solution will be similar for all the student groups. If the Eq.~\ref{eq:defq} is true, the solution $\mathcal{R}'$ will be of similar quality for all student groups.

\subsection{Flexibility of the formulation}
The FaiREO definition in Sect.~\ref{sec:faireo} can accommodate many different scenarios depending on what we consider the fair distribution of the courses, $\mathbf{X}$, to be. In every case, each row of $\mathbf{X}$ will sum up to one, i.e.,
$\sum_{p=1}^{g_s} x_{j,p} = 1, \quad \forall j \in \mathcal{C}.$

\begin{itemize}
    \item \textbf{Population-driven distribution.} As an initial starting point or in absence of insights about the desirable distribution of the course recommendations to student groups, we could set 
    \begin{equation}
        x_{j,p} = \frac{n_p}{n}, \quad \forall j \in \mathcal{C}, \quad\forall p \in \{1, \dots, g_s\}.
        \label{eq:popbased}
    \end{equation}
    
    That would set the fair distribution to match the underlying distribution of the population to the student groups. In this case, all the rows of the $\mathbf{X}$ matrix will be the same.
    \item \textbf{Coarse-grained distribution.} Instead of using the population distribution to define $x_{j,p}$, the department can decide what is the desirable distribution for all the courses. For example, if equally recommending a course to the student groups is not realistic, the department can assign an arbitrary fair recommendation ratio for all the courses in a department. This approach can potentially be less strict but more likely to be achieved in reality. It could assist the achievement of the department's (i.e., system's) goals regarding the diversity in course registration.
    \item \textbf{Fine-grained distribution.} In its most general form, we have the ability to assign different fair distributions for each course. While this formulation allow us to define different fair distributions for the courses, it would require detailed insights and considerable fine-tuning on the administrator's side.
    
\end{itemize}

\section{Related Work}
\label{sec:relwork}
In this section we explain how our problem relates to similar problems with respect to fairness. 
We have identified three main classes of problems closely related to our problem and fairness issues, and we present briefly representative work in the following subsections. As a side note, this paper refers to user group fairness for course recommendation, which is a different problem from group recommendation, where you aim to recommend the same set of items to every user in a group. In the problem we examine, we offer individual and personalized course recommendations to the users (in our case, the students) which need to be of as high quality as possible, while they are fairly distributed across the different student protected groups.

\subsection{Fair Recommender Systems}
Fairness in recommender systems is relatively new and each work presents its own point of view on the subject. 
A body of work studies fairness in ranking lists~\cite{zehlike2017fa, beutel2019fairness, biega2018equity, singh2018fairness, yang2017measuring, abdollahpouri2019managing, ge2021}, where the goal is to provide diverse representation of the items, i.e., equal exposure of different groups of items. A ranking is considered to be unfair when specific protected groups of items are under-ranked and as a result they receive lower visibility in the system. This corresponds to fairness with respect to items.
Beutel et al.,~\cite{beutel2019fairness} also account for user engagement.
They measure the differences in accuracy across the groups of items based on pairwise comparisons. 
According to their definition of pairwise fairness, assuming that two items have received the same user engagement, then both protected groups should have the same likelihood of a clicked item being ranked above another relevant unclicked item. 
%
Deldjoo et al.~\cite{deldjoo2021} proposed a generalized cross entropy measure of fairness that was based on a fair distribution of a model's performance over items or users. Yao et al.~\cite{yao2017} propose fairness metrics so that the error is fairly distributed across users.
The work in~\cite{zehlike2017fa} studies fairness in search engines of people, such as job recruiting, companionship, or friendship search. In such cases, an outcome is unfair if members of one protected group are systematically under-ranked than those of another protected group. The recommended candidates are determined by a ranking algorithm. The proposed method to remove the bias is a post-processing process. 
All the above works are different from ours as we do not account for diversity in the recommendation lists.
In the course recommendation domain, we recommend a limited number of courses.
Item diversity does not guarantee group fairness with respect to equality of opportunity for the student groups. 

The most relevant work is that of Marras et al.~\cite{marras2020}, which is also based on equality of learning opportunities in content recommendation. Different desirable properties of the recommended items and their measures are proposed, and the goal is for every list of recommendations to satisfy them above some threshold. While this approach is based on the same principle, their final outcome is different, as they focus on individual fairness.
Our goal in this paper is to recommend each course fairly across the user protected groups.
We aim at ensuring user-side fairness in a CRS and we consider protected groups of users/students, and not items. 

Another body of work studies fairness across multiple stakeholders in recommendations: the system, the suppliers/vendors/providers and the users/consumers~\cite{burke2016towards, mehrotra2018towards, burke2018, malthouse2019multistakeholder}. 
While these works study the trade-off between the different stakeholders' benefits, we are interested in user-side fairness only. We consider ways to improve fairness while harming as little as possible the relevance of recommended items, both of which are benefits for a single stakeholder that compete with each other. From these approaches, the most relevant notion is C-fairness~\cite{burke2018}. It considers the disparate impact of the recommendation on protected groups of consumers. The proposed method, a modification of the Sparse Linear Method (SLIM), does not directly balance the recommendation lists. Rather, it balances the neighborhoods based on which the suggestions are generated for all the users. While their evaluation metric, equity score, captures a similar notion as the first part of our FaiREO definition, it can be computed only per item group (or item/course in our case) and only for two protected groups. As a result, it cannot evaluate the overall fairness of a recommendation solution.

Apart from the algorithmic fairness, issues may also arise from biases in the input data which the recommender system amplifies~\cite{yang2017measuring, farnadi2018fairness}. Tsintzou et al.,~\cite{tsintzou2018bias} proposed a metric called bias disparity to measure the difference between the bias towards different movie genres in user profiles (input) and in resulted recommendations (output). A similar work proposed a group-based metric to compare the preference ratio in the input and output data (recommendation lists) and quantify the degree to which recommendation algorithms may propagate any biases~\cite{lin2019crank}. More recently, researchers have also studied representations that do not expose sensitive feature information in the user modeling process~\cite{wu2021}.



\subsection{Fair Resource Allocation}
In the Fair Resource Allocation or Fair Division problem, we want to fairly divide a resource or \textit{goods} to \textit{agents} with different preferences over the resource~\cite{bouveret2016fair}. In the course recommendation context, we could consider the courses as the resources, the students as the agents and the recommendation scores as the expressed preferences of the agents towards the goods.
Under this setting, group fairness has been studied in the form of envy-freeness~\cite{bouveret2016fair, berliant1992fair, husseinov2011theory}.
In an assignment, a group is treated fairly when each agent has no envy for the goods assigned to other agents; everyone gets what they value the most. In our case, this is already achieved by the HSC solution which is an envy-free solution. In order to ensure group fairness, we want to refine this initial solution and allocate/recommend different courses equally across the protected groups.

In certain works~\cite{manurangsi2017asymptotic, suksompong2018approximate}, the notion of group fairness deals with settings in which the members of each group are allocated the same set of resources, which does not apply in our case of protected groups, as each student of one group can receive different recommendations from the others in the same group. Aleksandrov et al., \cite{aleksandrov2018group} assume that each group has an aggregate preference for a specific bundle of goods of another group and they consider arithmetic-mean group preferences; a feature that does not apply in the present work.

\subsection{Fair Course Allocation}
According to the Course Allocation problem~\cite{diebold2014course}, we have a set of students with preferences to courses, a set of courses with preferences to students (priority orderings over the students from the course administrator), and each course has a specific predefined capacity; the goal is to allocate students to seats of courses. Course Allocation is an instance of the combinatorial assignment problem if we consider no preferences on the courses side (one-sided preferences)~\cite{budish2011combinatorial}. 
In this domain, a highly unfair outcome could lead to some students assigned to their most preferable courses and some other students assigned to their least ones or even to zero courses~\cite{budish2011combinatorial}. Diebold et al.,~\cite{diebold2014course} compared two stable matching algorithms to a first-come-first-serve approach, a mechanism used in many institutions. A matching is considered stable when there is no student-course pair, such that both prefer one another to their current assignment~\cite{diebold2017matching}. In a more recent work, Diebold et al.,~\cite{diebold2017matching} evaluated multiple matching mechanisms with real data in the context of course allocation with indifferences-ties in school preferences.
%
%
This notion of fairness corresponds to individual and not group fairness. Additionally, in the Fair Course Recommendation problem, there are no restrictions (such as the capacity of a course) other than 
trying to maintain the highest possible quality of the recommendation.

\section{Proposed Methods}
\label{sec:devalg}
		Fair course recommendation according to \algname{} is a multi-objective optimization problem, that simultaneously tries to satisfy both conditions of equal opportunity and quality, as described in Section~\ref{sec:faireo}. We define two different objective functions ($O$ and $Q$) to capture each condition, and then we linearly combine them into our overall objective function.
		
		\subsection{Opportunity Objective}
		We quantify the first condition of fairness by using the mismatch between the quantities of Eq.~\ref{eq:deff}, i.e., the distance of course $j$ from the fair ratio that corresponds to protected group $p$, $x_{j,p}$. 
		We compute the \textit{fraction of the recommendations that introduce unfairness} in protected group $p$ as:
		\begin{equation}
        o_p = \frac{1}{n_{p} k} \sum_{j=1}^{m} \left(n^{(j)} \bigg\lvert  \frac{n^{(j)}_{p}}{n^{(j)}} - x_{j,p} \bigg\rvert\right),
		\label{eq:objFp}
		\end{equation}
        where $n$ is the number of students, $m$ is the number of courses, $n_p$ is the number of students belonging in group $p$, $k$ is the number of courses we recommend to the student, $n^{(j)}$ is the number of students to whom we recommend course $j$, and $n^{(j)}_p$ is the number of students from group $p$ to whom we recommend course $j$.
		The term of the absolute difference captures how far away we are from balancing the opportunities offered in group $p$ regarding course $j$. 
        The term in the parenthesis corresponds to the number of students that introduce this unbalance in the recommendations of $j$ to group $p$.
		Note that the overall sum is normalized with the number of recommendations generated for group $p$ ($k$ courses for every one of the $n_p$ students), in order for $o_p$ to be invariant of the group size.
        The opportunity objective function targets to minimize the unfairness existing in the recommended lists of courses:
        \begin{equation}
        O = \min \norm{\mathbf{o}}_l, \text{ where } \mathbf{o}= [o_1,\ldots,o_{g_s}].
		\label{eq:objF}
		\end{equation}
		The overall opportunity objective is measured by the $l$-norm of the vector $\mathbf{o}$.

		\subsection{Quality Objective}
		To quantify the quality objective, we use the recommendation scores of the courses. We measure the quality of any assignment of courses to students by the summation of the recommendation scores of the suggested courses. We will capture how different is the quality of a solution compared to the solution that recommends the highest scored courses to the students (HSC solution). We formulate the \textit{fraction of quality loss} for each group $p$ as:
		\begin{equation}
			q_p = \frac{\sum_{i \in \mathcal{S}_p} \big( \sum_{j \in \mathcal{R}_i} y_{i,j} - \sum_{j \in \mathcal{R}'_i} y_{i,j} \big) }
			{\sum_{i \in \mathcal{S}_p} \sum_{j \in \mathcal{R}_i} y_{i,j}},
		\label{eq:objQp}
		\end{equation}
		where $\mathcal{R}'_i$ is a set of recommended courses for student $i$, and $\mathcal{R}_i$ the set of courses recommended based on the HSC solution.
		$y_{i,j}$ is the recommendation score of student $i$ in course $j$, and $\mathcal{S}_p$ is the subset of students in group $p$. The summation in the numerator is the difference in quality between the two solutions. We normalize it to make it invariant of the size of the protected groups and their quality. 
		The quality objective function that minimizes the quality loss is:
		\begin{equation}
        Q = \min \norm{\mathbf{q}}_l, \text{ where } \mathbf{q}= [q_1,\ldots,q_{g_s}].
		\label{eq:objQ}
		\end{equation}
        The overall quality objective is measured by the $l$-norm of the vector $\mathbf{q}$.
        
        \subsection{Combined Objective Function}
        There is a trade-off between the two objectives, as optimizing for the opportunity objective will replace the highest-scored courses with others that have the same or lower scores. This may result in recommendations with lower-scored courses than the HSC solution, which will incur quality loss. The combined objective function is:
		\begin{equation}
            V = a O + (1-a) Q.
			\label{eq:objV}
		\end{equation}
		The parameter $\alpha \in [0,1]$ weighs the importance of each objective. 
		Note that when $\alpha$ takes a marginal value ($0$ or $1$), all the weight is placed in one objective ($Q$ or $O$). Thus, in these cases, it is very likely that unfairness will be introduced from the other unpenalized objective ($O$ or $Q$, respectively).
		We can use any $l$-norm greater than $1$, which penalize high values, to aggregate the vectors $\mathbf{o}$ and $\mathbf{q}$ for all student groups. 

	\subsection{Greedy Hill Climbing (GHC) Algorithms}
	\label{sec:devalg:ghcrm}
    The fair course recommendation is a multi-objective, combinatorial optimization problem described by Eq.~\ref{eq:objV}, and it involves a discrete but large configuration space. That space cannot be exhaustively searched, as there are $ \genfrac(){0pt}{1}{m}{k}^n$ possible combinations to examine, where $n, m$, and $k$ are the number of students, courses, and recommended courses per student, respectively.

    Our approach described in Alg.~\ref{alg:GHC} uses the steepest ascent hill climbing technique, with a greedy strategy for performing local search. It includes two phases: 1) the assignment of an initial solution, and 2) the refinement of this solution in order to reach a solution that better minimizes the objective function, $V$. The refinement consists of a series of moves that the algorithm makes towards a fairer solution. At every step, it performs a single change in one student's recommendation list by replacing a single course.  
    Iteratively, it considers a neighborhood of solutions that it can reach by making a single move from the current solution, and greedily selects the move that minimizes $V$. The algorithm terminates when it cannot find a single move that improves $V$. It will reach one local minimum out of many that might exist in such a combinatorial optimization problem. Additional details about these steps are provided in the subsequent sections.
    
    \begin{algorithm}[bt] 
		\caption{Greedy Hill Climbing (GHC)}
		\label{alg:GHC}
		\begin{algorithmic}[1]
			\Require $\mathbf{R}$ (Recommended courses for every student.)
			\Require $\alpha$ (Weight of the opportunity objective, needed to compute the $V$ objective.)
			\Require $\mathcal{M}_{\mathbf{R}}$ (Allowed moves we can make from solution $\mathbf{R}$ according to GHC(None) or GHC(GC).)
			\State $V \gets$ Objective value of solution $\mathbf{R}$. 
            \State $\mathbf{R}' \gets \mathbf{R}$ \Comment{Current solution.}
            \State $\mathcal{M}_{\mathbf{R}'} \gets \{(i,j_\text{out}, j_\text{in})\}$ \Comment{Set of possible moves to reach neighboring solutions.}
			\State $\vt \gets \emptyset, \vT \gets \emptyset$ \Comment{Visited course and student groups while not finding an improved solution.}
			\While{$\lvert\vt\rvert<m$}
				\State $ V' \gets$ Objective value of solution $\mathbf{R}'$. 
				\For{$(i,j_\text{out}, j_\text{in}) \in \mathcal{M}_{\mathbf{R}'} $} \Comment{Possible moves.}
				
				\State $\mathbf{R}'' \gets \mathbf{R}'$
				\State $\mathcal{R}''_i \gets \mathcal{R}''_i - \{j_\text{out}\} + \{j_\text{in}\} $
				\State $ V'' \gets$ Objective value of solution $\mathbf{R}''$.
	      		\If {$V'' \leq V'$} \Comment{Store the best move.}
					\State $V' \gets V''$
					\State $(i',j'_\text{out}, j'_\text{in}) \gets (i,j_\text{out}, j_\text{in})$
				\EndIf
				\EndFor
				\If {$V' \leq V$} \Comment{Make a move.}
				\State $\mathcal{R}'_{i'} \gets \mathcal{R}'_{i'} - \{j'_\text{out}\} + \{j'_\text{in}\} $
				\State Update $\mathcal{M}_{\mathbf{R}'}$.
				\State $V \gets V'$
				\State $\vt \gets \emptyset, \vT \gets \emptyset $ 
				\Else
				\State Update $\vt, \vT$. \Comment{Add the examined target course/student group in the visited list. }
				\EndIf
			\EndWhile
			\State \textbf{return} $\mathbf{R}'$
		\end{algorithmic}
	\end{algorithm}
    
    \subsubsection{Initial solution}

    We first need to decide which will be the initial solution for our refinement algorithm.
    A common practice is to start from a good solution and try to improve it. There is one solution that minimizes the quality objective; that is the HSC solution. On the other hand, there are many solutions that minimize the opportunity objective without considering the value of recommending a particular course to a student. Since we have access to the recommendation scores of a CRS model (assumption 2, Sect.~\ref{sec:assumpt}), we can use HSC solution as the initial assignment. By design, HSC achieves $Q=0$, which is the global minimum w.r.t. the $Q$ objective.
    We start from the HSC assignment and refine it to support the notion of \algname{} fairness.
    
    \subsubsection{Moves}
    
    A fundamental element of search methods is the type of moves allowed to transition from a feasible solution to another one. Given a solution, we remove a course from the recommendation list of a single student, and replace it with another course. This move can be fully described by a triplet $(i, j_\text{out}, j_\text{in})$, where $j_\text{out}$ and $j_\text{in}$ are the courses we remove from, and introduce to the recommendation list of student $i$, respectively.
    A move is \textit{positive} when it results in a solution with lower objective function $V$, and \textit{negative}, otherwise.
    
    \subsubsection{Neighborhood of solutions}
    \label{sec:devalg:ghcrm:neighbor}
    Assuming that we recommend courses to students based on the solution $\mathbf{R}'$, we need to specify the neighborhood of solutions that the algorithm will evaluate in order to make a move that improves the combined objective function, $V$, the most. The corresponding set of candidate moves is denoted by 
    $\mathcal{M}_{\mathbf{R}'}$. We study two different ways to define them by specifying the allowed values for $(i, j_\text{out})$, which result in two algorithms. Both of them follow the steps of the Alg.~\ref{alg:GHC}, but examine different set of moves $(i, j_\text{out}, j_\text{in})$, $\mathcal{M}_{\mathbf{R}'}$. Given a pair of student and recommended course in $\mathcal{R}'_i$, $(i, j_\text{out})$, we examine all the courses not currently recommended to the student $i$ as candidate courses, i.e., $j_{\text{in}} \in \mathcal{C} - \mathcal{R}'_i$, to complete the set of moves in $\mathcal{M}_{\mathbf{R}'}$.

    The simplest solution is to examine all possible one-step-away solutions from the existing solution. This is a full-blown search that will consider changing all student-course pairs $(i, j_\text{out}), \text{ where } i \in \mathcal{S}\text{, and }j_\text{out} \in \mathcal{R}'_i$. In this method, \textbf{GHC(NoNe)}, there is essentially no neighborhood specified. If the algorithm examines all moves in $\mathcal{M}_{\mathbf{R}'}$ and cannot find a positive one that improves the $V$ objective, it terminates. In this case, the Algorithm~\ref{alg:GHC} will reach lines 19--20, and $\vt, \vT$ will include all the courses and student groups, respectively. As a result, in the next iteration, the while loop in line 5 will be false and the algorithm will terminate.
    
    We also use some heuristics for defining a neighborhood in order to avoid searching all the space every time. Rather, we examine a smaller set of moves, hoping that the next best move will belong there. In the \textbf{GHC(Gc)} method, we consider moves altering only the recommendations of students in a specific (target) protected group $T$, $i \in \mathcal{S}_T$, for a specific (target) course $t$, $j_\text{out}=t$. We choose the \textit{target protected group} $T$ to be the one that exhibits the highest opportunity unfairness, i.e., the most severe unbalance in recommendations:
    $$T = \underset{p \in \{1, \dots, g_s\}-\mathcal{V}_T}{\text{arg}\max} o_p,$$
    where $\mathcal{V}_T$ is the set of student groups we have already visited and considered that do not result in positive move.
    The \textit{target course} $t$ is the one that is over-recommended the most among the students of group $T$, i.e., 
    $$t=\underset{j \in \mathcal{C} - \mathcal{V}_t}{\text{arg}\max\text{ }} o_{p,j} \text{, \quad where }o_{p,j} = n^{(j)} \left( \frac{n^{(j)}_{p}}{n^{(j)}} - x_{j, p} \right),$$
    where $\mathcal{V}_t$ is the set of courses we have already visited that do not result in positive move.
    $o_{p,j}$ is the term in the parenthesis in Eq.~\ref{eq:objFp} without the absolute value.
    We select $(i, j_\text{out})$ in such a way assuming that the student group $T$ and the course $t$ have the most room for improvement during refinement.
    If the algorithm cannot find a better solution, it adds $t$ to the set of visited courses $\mathcal{V}_t$ (Alg.~\ref{alg:GHC}, line 20) and finds the next target course to search. Once there are no courses left to consider as target courses for $T$, we empty $\mathcal{V}_t$, add $T$ to $\mathcal{V}_T$ (Alg.~\ref{alg:GHC}, line 20), and explore the next target student group. The algorithm terminates when we have visited all the student groups and courses but cannot find a positive move.

    \subsubsection{Computational complexity of a move}
    In terms of computational complexity, with GHC(NoNe), we need to examine the whole search space every time we make a move, which includes $n k (m-k)$ solutions. This reflects the fact that we need to consider changing each recommendation ($n k$) with every course not currently recommended to the student ($m-k$).
    To perform one move with GHC(Gc), we need to find the target course $t$ and protected group $T$, which entails examining $n_g+m$ elements in the worst case. Then, we need to evaluate the moves within the specified neighborhood which involves changing each recommendation of course $t$ in students of the group $T$ if $t \in \mathcal{R}'_i$ for $i \in T$, which includes $n^{(t)}_{T}$ instances. In total, the complexity of the GHC(Gc) algorithm is $n_g+m+n^{(t)}_{T}(m-k)$. 

    \subsection{Incremental GHC Algorithm (GHC-Inc)}
\begin{figure}[tb]
        \centering
        \includegraphics[width=0.5\linewidth]{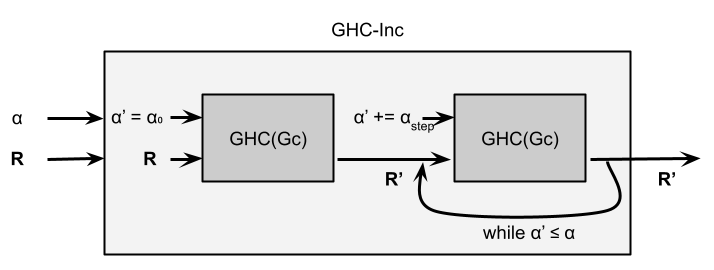}
        \caption{A diagram of the GHC-Inc method, which uses the GHC(Gc) algorithm as its components. Both GHC-Inc and GHC(Gc) receive as input a value of $\alpha$, which weighs the importance between the opportunity and quality objectives, and a recommendation solution from a fairness-unaware model, $\mathbf{R}$. Their output is an updated recommendation solution, $\mathbf{R}'$. GHC-Inc iteratively uses GHC(Gc), where it starts from an initial value $\alpha_0$ and gradually increases its value by $\alpha_\text{step}$. The output of GHC(Gc) at each step is fed to the next model of GHC(Gc) as its initial recommendation solution.} 
        \label{fig:ghc-inc}
    \end{figure}

    We also propose another algorithm to optimize the overall objective function, based on GHC(Gc). In the \textbf{GHC-Inc} algorithm, instead of optimizing for the given parameter $\alpha$, we start optimizing the objective function with a small value of $\alpha' \gets \alpha_0$. Once we reach a local minimum, we increment alpha by a parameter $\alpha_\text{step}$, and we further improve the current solution for the updated value of $\alpha' \gets \alpha'+\alpha_\text{step}$. We repeat this until we reach the value of given parameter $\alpha' = \alpha$, as show in Figure~\ref{fig:ghc-inc}. We \textit{gradually increase} the importance of the opportunity objective, in order to take careful steps in the beginning that do not introduce a high quality loss. Our goal is to reach a more balanced assignment with a lower $Q$ objective.
    
    \subsection{Tabu-based GHC Algorithm (GHC-Tabu)}
    \label{sec:devalg:tabu}
    So far, the discussed algorithms stop exploring the solution space when they have reached a local minimum, i.e., there is no single move that would improve the current solution. However, in such a huge solution space, this might not be the global minimum. To further explore the search space after this point, we incorporate the idea of Tabu search in the GHC(Gc) algorithm. Whenever there are no improving moves and GHC(Gc) would stop, the \textbf{GHC-Tabu} algorithm performs the move that degrades the objective function the least. We hope that by taking a negative move, we will get into a different neighborhood of solutions that will drive us to a better local minimum. In order for the algorithm to terminate, we control the \textit{number of negative moves} that we allow it to make.
    
    We also need to ensure that the algorithm will not make the reverse move on the next step, and return to the local minimum already visited. We use the tabu list, a short-term memory list structure with the first-in-first-out property, to store every move we made in order not to reverse it. A parameter controls the \textit{tabu list size}. We store the pair of student-course $(i,j_\text{in})$ that we just updated and that we do not allow to take back. 
    However, reversing a move can sometimes lead to a better solution. We introduce an \textit{aspiration criterion} which allows us to make moves forbidden by the tabu list if they lead us to a solution with lower objective function than the lowest objective achieved so far.

\section{Experimental Setup}
\label{sec:setup}
    In Sect.\ref{sec:results}, we will present the experimental results when we use as fair distribution the population-based distribution. In this case, the fair ratio $x_{j,p}$ for each course $j$ is expressed by the Eq.~\ref{eq:popbased}.
	\subsection{Synthetic Datasets}
		
		We generated synthetic datasets to evaluate our approaches since we do not have data regarding the students' protected attributes. The kind of data that that we need to generate are: 
		\begin{enumerate*}
		    \item the student-course recommendation score matrix $\mathbf{Y} \in \mathds{R}^{n \times m}$, where $y_{i,j}$ represents the recommendation score of course $j$ for student $i$ estimated by any CRS, and
		    \item the partitioning of students into protected groups, $\mathcal{S}_1,\dots,\mathcal{S}_{g_s}$. 
		\end{enumerate*}
		We want to create synthetic datasets whose characteristics align with these of real-world datasets. Let us assume a matrix $\mathbf{Y}$ obtained from a CRS and the corresponding solution when we recommend the highest scored courses for each student. The fairness of this solution depends on the existence of courses whose recommendation scores tend to be higher for a specific group of students. In that case, these courses will be good candidates and systematically suggested more times to one particular group than the rest. We model these factors into our dataset generator, by introducing the notion of \textit{course buckets}, which correspond to a partition of the set of courses $\mathcal{C}$. The number of course buckets is controlled by the parameter $g_c$. Each bucket of courses will have different average recommendation score across the protected groups. 
		
		
		We model the relation between student groups and course buckets via a matrix $\mathbf{M} \in \mathds{R}^{g_s \times g_c}$, such that $m_{p,q}$ is the average value of the recommendation scores of students in group $p$ for courses in bucket $q$. We fill the first row of $\mathbf{M}$, $\mathbf{M}_0$, by sampling a normal distribution $N(\mu_\text{M}, d_\text{M})$ with mean value $\mu_\text{M}$, and standard deviation $d_\text{M}$. We fill the remaining rows of $\mathbf{M}$ with a permutation of the initial vector $\mathbf{M}_0$. This ensures that all students will have some high-scored courses, and the recommendation quality across student groups will be similar. Once we have generated matrix $\mathbf{M}$, we can finally fill the matrix $\mathbf{Y}$ by sampling the recommendation scores for students in group $p$ and courses in bucket $q$ from a normal distribution $N(\mu_\text{Y} = m_{p,q}, d_\text{Y})$. 
		All in all, in order to generate a dataset based on this process, we need to specify the following parameters: number of course buckets $g_c$, $\mu_\text{M}, d_\text{M}$ for the initial vector of means $\mathbf{M}_0$, and the standard deviation $d_\text{Y}$ for the generation of $\mathbf{Y}$.
		
        \begin{figure}[b]
            \centering
            \includegraphics[width=0.55\linewidth]{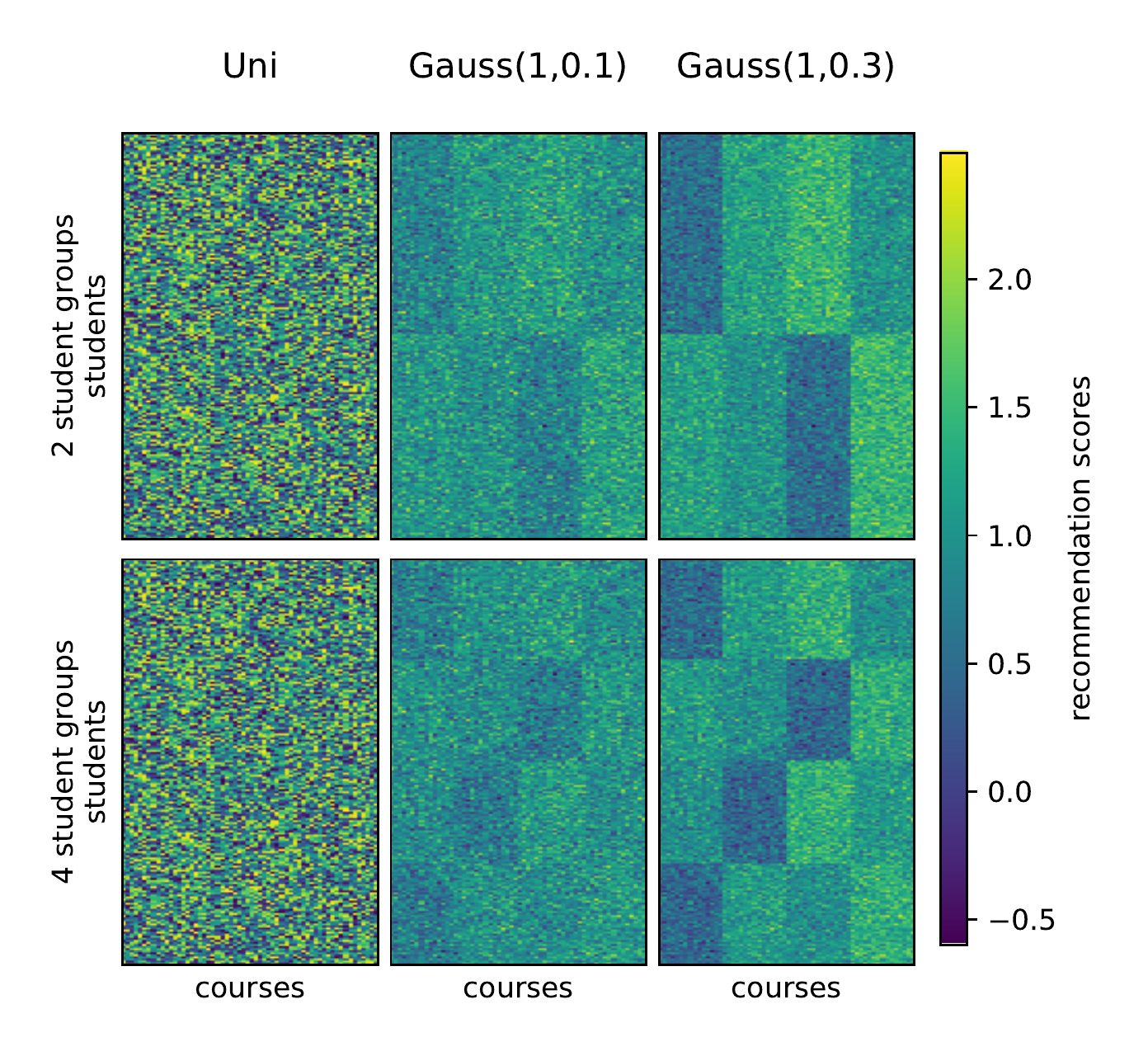}
            \caption{Heat maps of all the students' recommendation scores for all courses in each synthetic dataset.}
        	\label{fig:plotHeatmap}
            
        \end{figure}
        
		This dataset generator is parameterized in order to create datasets of different difficulty levels. This allows us to evaluate how our algorithm will operate under different settings. The difficulty of a dataset is controlled by how close to each other are the mean values in $\mathbf{M}_0$, i.e., the standard deviation $d_\text{M}$. The further away the means are, the further away the scores of different protected groups for a course bucket will be (and the less likely it will be to recommend courses from this bucket to all student groups).
		It affects how many high scores exist for every student in the resulting dataset. If there are many courses with high scores for a student, then it will be more likely to find a move that improves the opportunity objective without introducing a high quality loss.
%

        We set the parameters as follows: $n=600, m=60, g_s=\{2,4\}, g_c=4, d_\text{Y}=0.3$. We generated datasets of three difficulty levels: easy (\texttt{Uni}), medium (\texttt{Gauss(1, 0.1)}), and hard (\texttt{Gauss(1, 0.3)}). 
        For the datasets \texttt{Gauss(1, 0.1)} and \texttt{Gauss(1, 0.3)}, we set $\mu_\text{M}=1.0$ and $d_\text{M}=\{0.1, 0.3\}$, respectively.
        In \texttt{Gauss(1.0, 0.1)}, the means generated in $\mathbf{M}$ will be closer to each other compared to \texttt{Gauss(1.0, 0.3)} datasets.
        When the means are spread out in a wider range in the matrix $\mathbf{M}$, the recommendation scores generated based on that matrix will have different statistical characteristics. The \texttt{Gauss(1.0, 0.3)} datasets will be harder datasets to handle, and we expect to incur a higher quality objective value. 
        The easiest datasets are \texttt{Uni}, where all recommendation scores $y_{i,j}$ are sampled from a uniform distribution in $[0,1]$, and there are not courses with high scores by design.
        In total, we create \textbf{six families of datasets}; for three difficulty levels, and for two or four protected groups. For every family of datasets, we create five versions of them, by using different seeds to generate the matrices $\mathbf{M}$ and $\mathbf{Y}$. In this way, we get to examine how sensitive are our models to input data with similar characteristics.  Figure~\ref{fig:plotHeatmap} shows the heatmap of the recommendation scores for each student ($y$-axis) and course ($x$-axis) in example datasets of each family. 


	\subsection{Real Datasets}
	We collected data from the Computer Science and Engineering department in the University of Minnesota. The data include the grades of undergraduate students and span a period of 10 years, until the fall semester of 2015. We only considered full-time students that actually graduated with a bachelor's degree. We used the last three semesters to test a recommendation system~\cite{polyzou2019} which was built using the rest of the data. We keep the recommendation scores of the students in the test set, and use them as the matrix $\mathbf{Y}$. We treat every semester as a different dataset: fall 2014, spring 2015 and fall 2015, with $188, 170, 112$ students and $59, 54, 55$ courses, respectively. We will be referring to these datasets as the \texttt{ComptSci} datasets.
	 
    The available data did not include any protected attributes that we could consider for our experimental evaluation, so we used other student-related information (entry registration status and the number of credits transferred) to \textit{simulate the socioeconomic status of the students}. 
    This led to three protected groups: high school students with less than $15$ credits transferred (HS), high school students with more than $15$ credits transferred (HSAP), and those coming from other institutions/colleges (NAS). 
    High school students can take Advanced Placement (AP) courses and transfer the credits earned to their undergraduate program.
    Minorities and low-income students are underrepresented in AP classes, and a low percentage of them actually take and pass the AP exams~\cite{tugend2017benefits,whiting2009multi}.
    Regarding NAS students, we do not have information about the institution where they transferred from. However, reports statistically show that almost half of NAS students come from 2-year colleges~\cite{shapirotracking} which are considered a major access point to 4-year institutions for minority and low-income students~\cite{crisp2014understanding}. The fraction of students in our datasets belonging in the HS, HSAP, and NAS protected groups is shown in Table~\ref{tab:group_distr}.

    \begin{table}[tb]
        \centering
        \caption{Statistics of the \texttt{ComptSci} datasets regarding the total number of students and their distribution over the three student groups we considered.}
        \label{tab:group_distr}
        \begin{tabular}{l|rrr|l}
        \toprule
             & Fall 2014 & Spring 2015 & Fall 2015 & Common Student Characteristics \\
        \midrule
        $n$  & 188       & 170         & 112       &                                \\
        HS   & 0.19      & 0.18        & 0.21      & Minorities, low-income         \\
        HSAP & 0.76      & 0.76        & 0.73      & High-income                    \\
        NAS  & 0.06      & 0.05        & 0.06      & Low-income \\ \bottomrule
\end{tabular}
    \end{table}
    
	\subsection{Model Parameters}

		Regarding the norm base, $l$, we use the $L_\infty$ norm.
		This norm considers only the highest elements of the vectors $\mathbf{o}$ and $\mathbf{q}$. $L_\infty$ is a strict norm, that forces the student group with the highest objective values to get as low as possible. This will limit the worst case scenario for the protected groups. 
		
        We need to specify the number of courses to recommend $k$, and the value of $\alpha$ that controls the trade off between opportunity and quality loss objectives. We set $k=5$, and $\alpha=\{0.1, 0.2, 0.3, 0.4, 0.5, 0.6, 0.7, 0.8, 0.9\}$. If $\alpha=0$, we get the initial, fairness-unaware recommendation, HSC solution. If $\alpha=1$, we would get a recommendation solution that does not consider at all the recommendation scores, and it could end up being worse than random assignment in terms of quality.
        
        For the algorithm GHC-Inc, we set $\alpha_0=0.1, \alpha_\text{step}=0.1$. 
        For the GHC-Tabu, we set the number of negative moves to $150$ and the tabu list size to $50$.

\section{Experimental Results}
\label{sec:results}
\begin{figure}[bt]
	 	\centering
	 	\begin{subfigure}[b]{0.7\linewidth}
	 	    \centering
            \includegraphics[width=\linewidth]{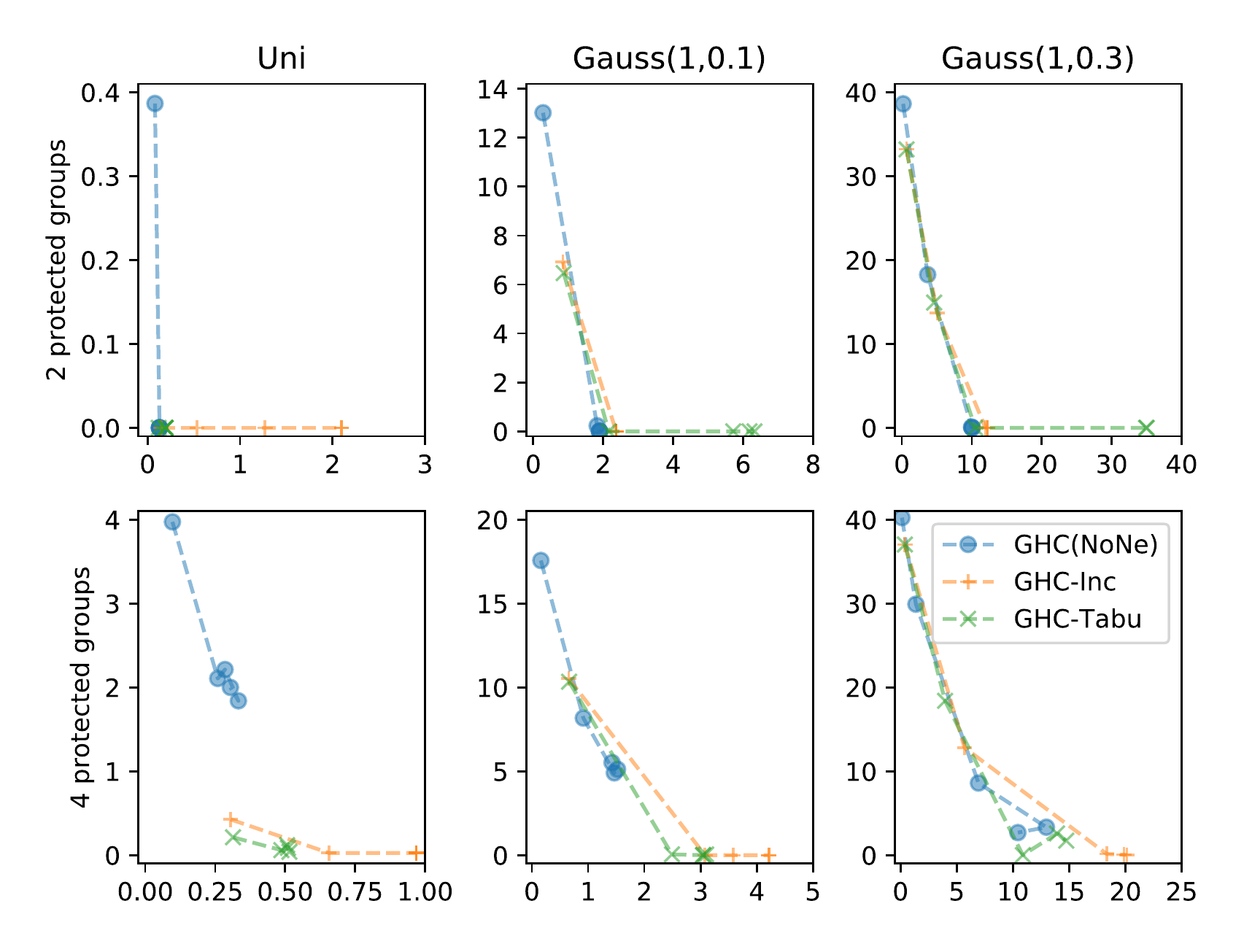}
		 	\caption{Synthetic datasets. Datasets in a row have increasing difficulty.}
		 	\label{fig:plotFQObjV-Synth}
		 	\vspace{1em}
        \end{subfigure}
        \begin{subfigure}[b]{0.7\linewidth}
	 	    \centering
		 	\includegraphics[width=\linewidth]{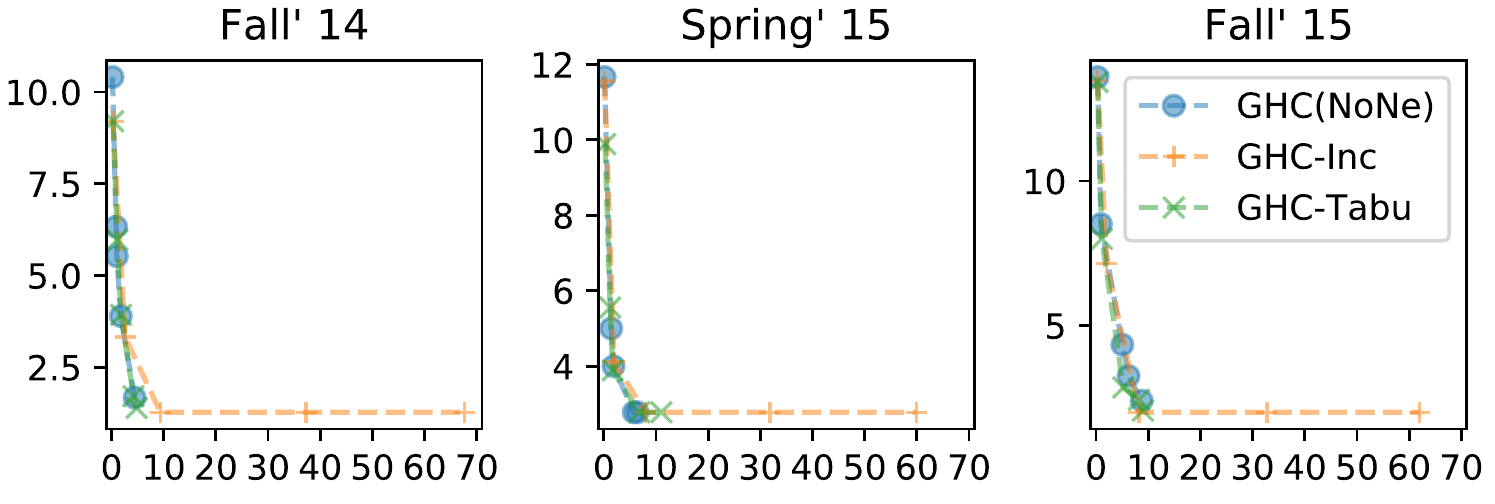}
		 	\caption{\texttt{ComptSci} datasets.}
		 	\label{fig:plotFQObjV-Compt}
		\end{subfigure}
		\caption{Scatter plots of the opportunity and quality objectives for the synthetic and \texttt{ComptSci} datasets with $\alpha = \{0.1,0.3,0.5,0.7,0.9\}$. For each dataset, the scatter plot shows the objective values $Q$ and $O$ achieved for different values of $\alpha$. A line connects points that correspond to consequent values of $\alpha$. The $x$ axis is the quality degradation percentage (\%) of the most degraded group and the $y$ axis is the percentage (\%) of unfair recommendations of the most impacted group w.r.t. opportunity.
		These two quantities are computed as  $100\times Q$ and  $100\times O$, respectively. 
        The $(0,0)$ point represents the ideal model that is fair for all student groups in terms of both quality and opportunity. 
		}
		\label{fig:plotFQObjV}
        
 \end{figure}

\begin{figure}[tb]
        \centering
     	\begin{subfigure}[b]{0.5\linewidth}
     	    \centering
            \includegraphics[width=\linewidth]{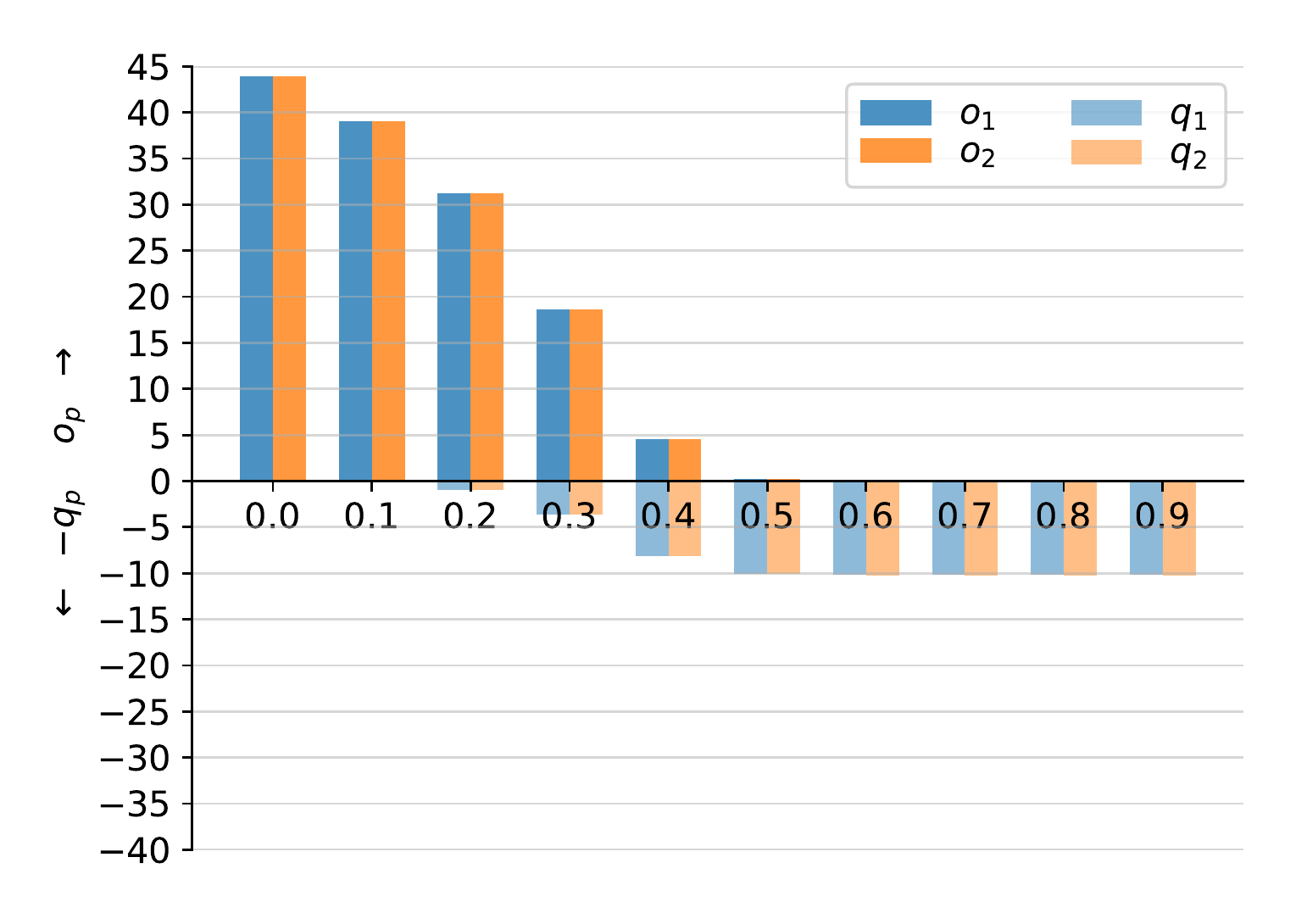}
            \caption{GHC(NoNe)}
            \label{fig:objDistrAlpha2a}
        \end{subfigure}%
        \begin{subfigure}[b]{0.5\linewidth}
     	    \centering
            \includegraphics[width=\linewidth]{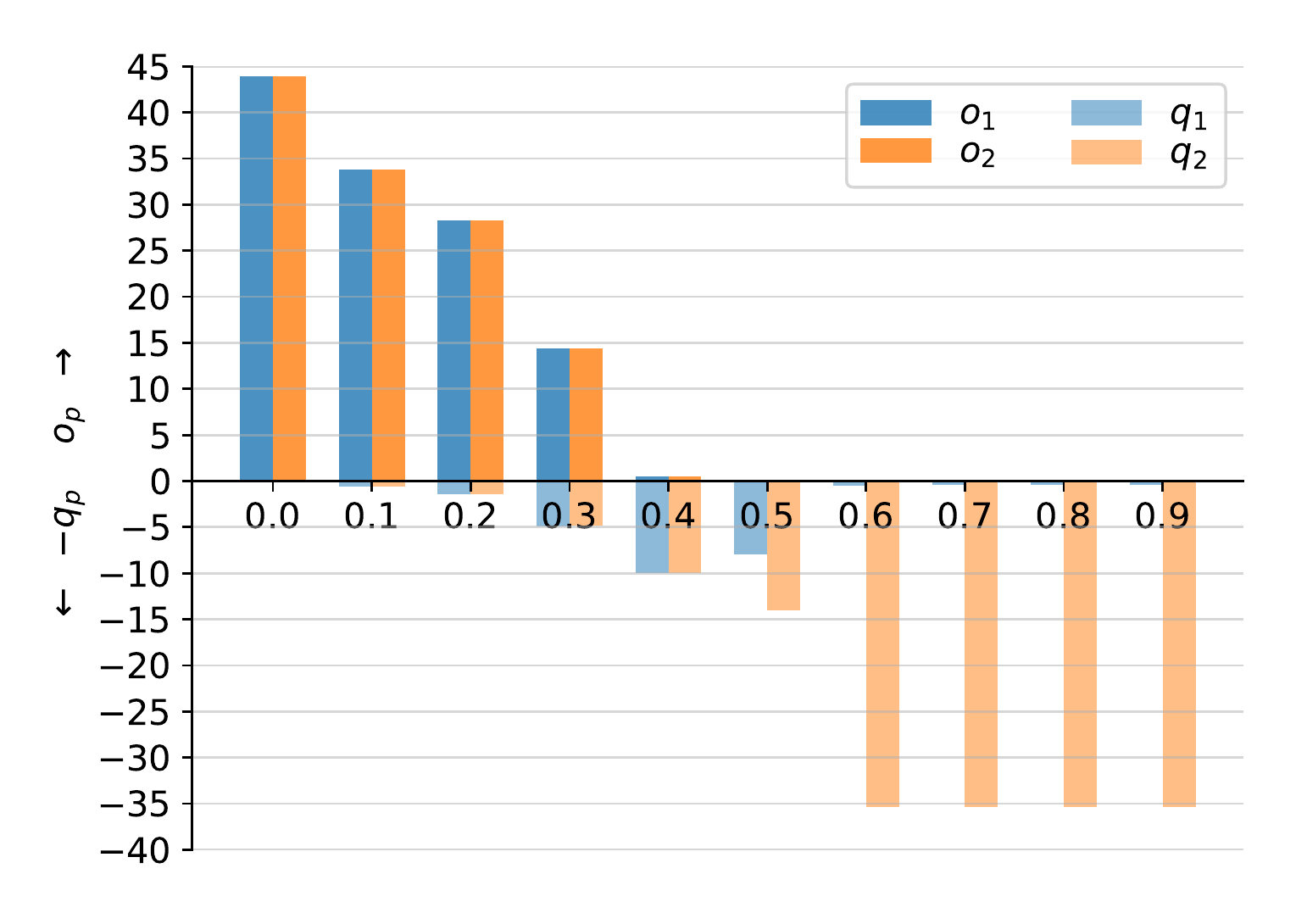}
            \caption{GHC(Gc)}
            \label{fig:objDistrAlpha2b}
        \end{subfigure}
    
        \begin{subfigure}[b]{0.5\linewidth}
     	    \centering
    	 	\includegraphics[width=\linewidth]{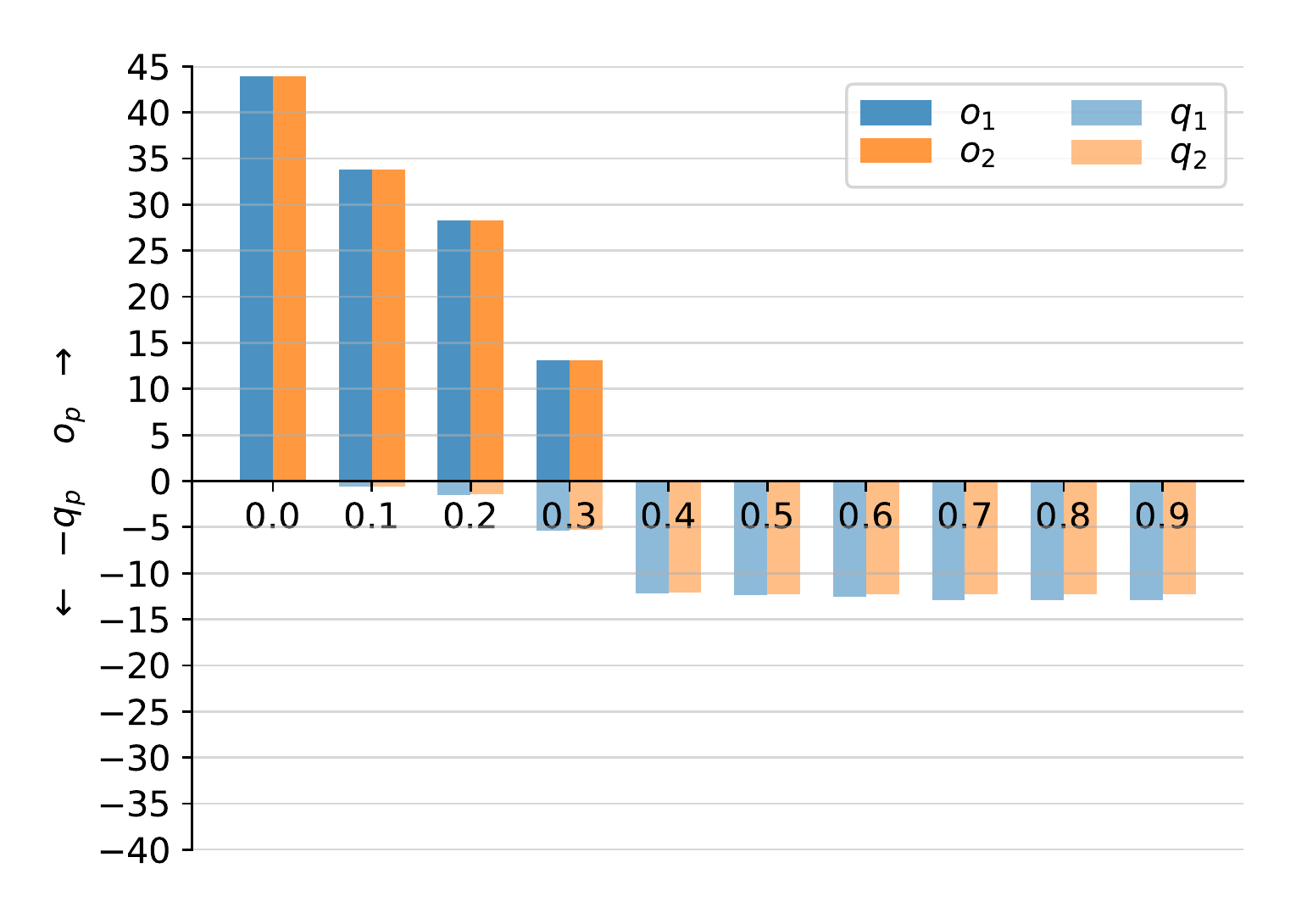}
            \caption{GHC-Inc}
            \label{fig:objDistrAlpha2c}
    	\end{subfigure}%
    	\begin{subfigure}[b]{0.5\linewidth}
     	    \centering
    	 	\includegraphics[width=\linewidth]{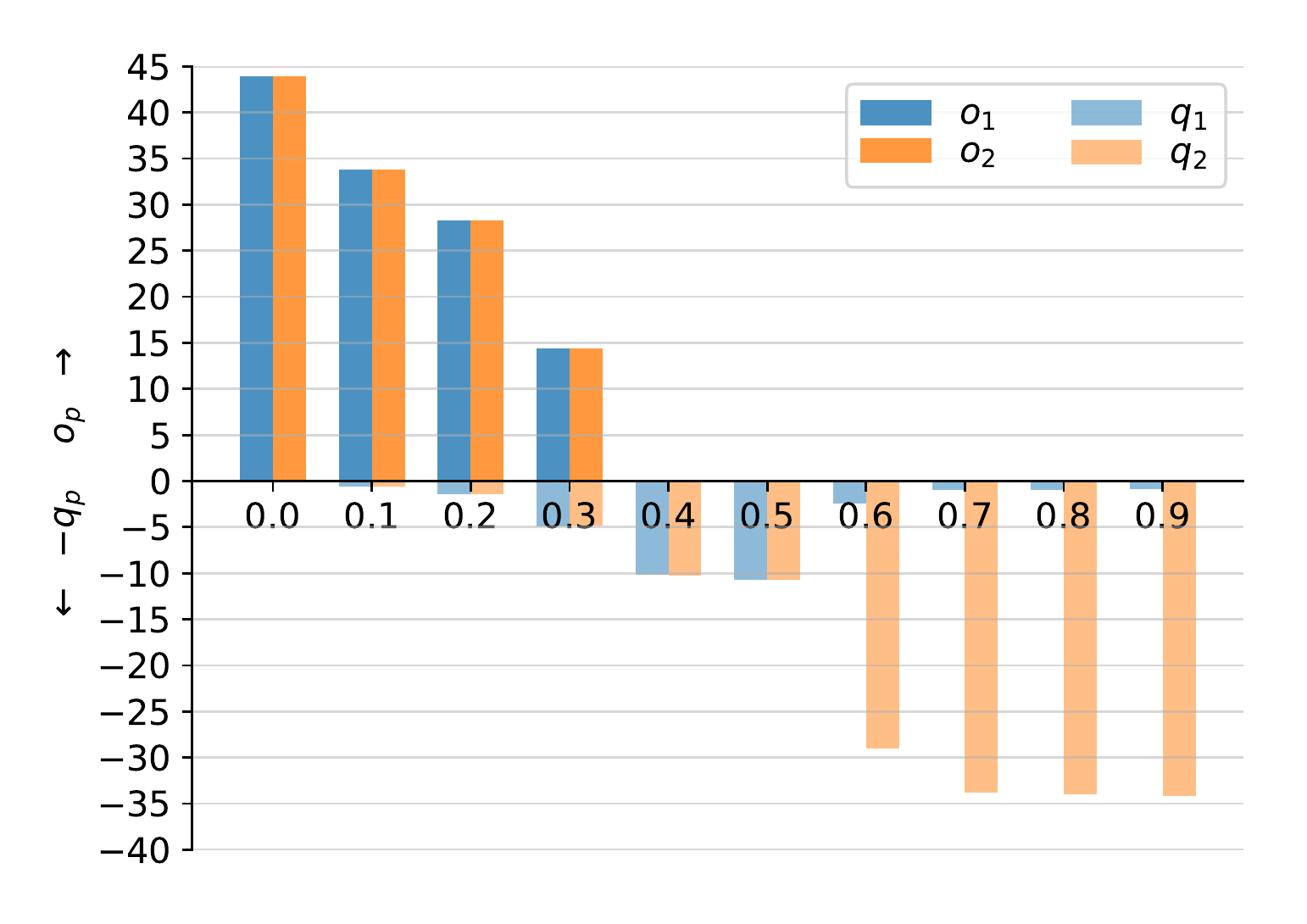}
            \caption{GHC-Tabu}
            \label{fig:objDistrAlpha2d}
    	\end{subfigure}
    
    	\caption{Distribution of the opportunity and quality objective values for different values of $\alpha$ in the difficult synthetic dataset \texttt{Gauss(1,0.3)} with two student groups for the four different methods. The values on the $y$-axis are multiplied by $100$ to correspond to percentages.
    	The $x$-axis represents different values of $\alpha$.
    	The positive side of the $y$-axis is the percentage of unfair recommendations w.r.t. opportunity of a student group, while the negative side shows the percentage of quality loss incurred for a student group (multiplied by -$1$). We also include the values for $\alpha=0$, which correspond to the initial values of the objectives achieved by the HSC solution, with high opportunity objective but zero quality objective.
    	}
    	\label{fig:objDistrAlpha2}
            
    \end{figure}

\begin{figure}[tb]
        \centering
     	\begin{subfigure}[b]{0.5\linewidth}
     	    \centering
            \includegraphics[width=\linewidth]{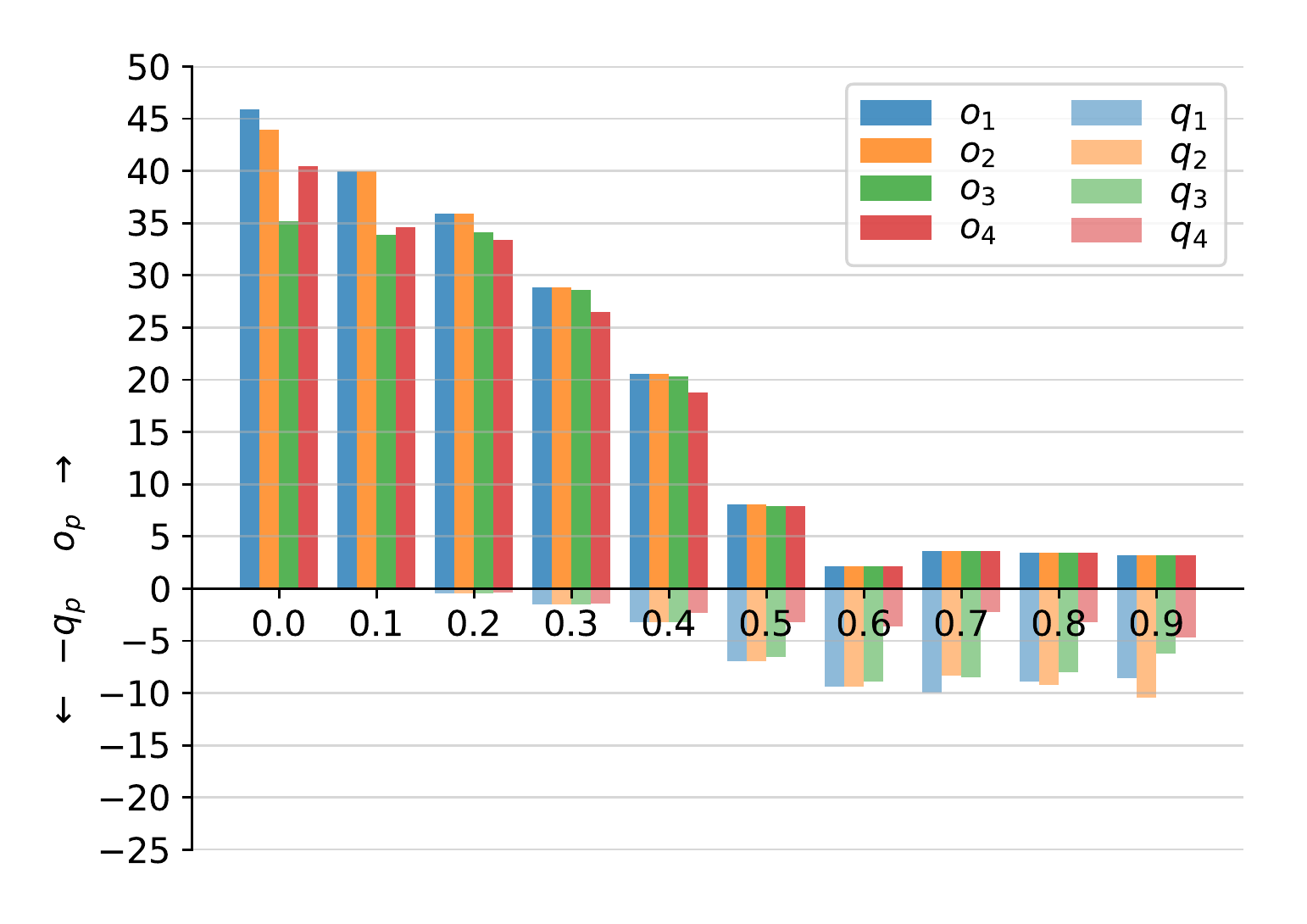}
            \caption{GHC(NoNe)}
            \label{fig:objDistrAlpha4a}
        \end{subfigure}%
        \begin{subfigure}[b]{0.5\linewidth}
     	    \centering
            \includegraphics[width=\linewidth]{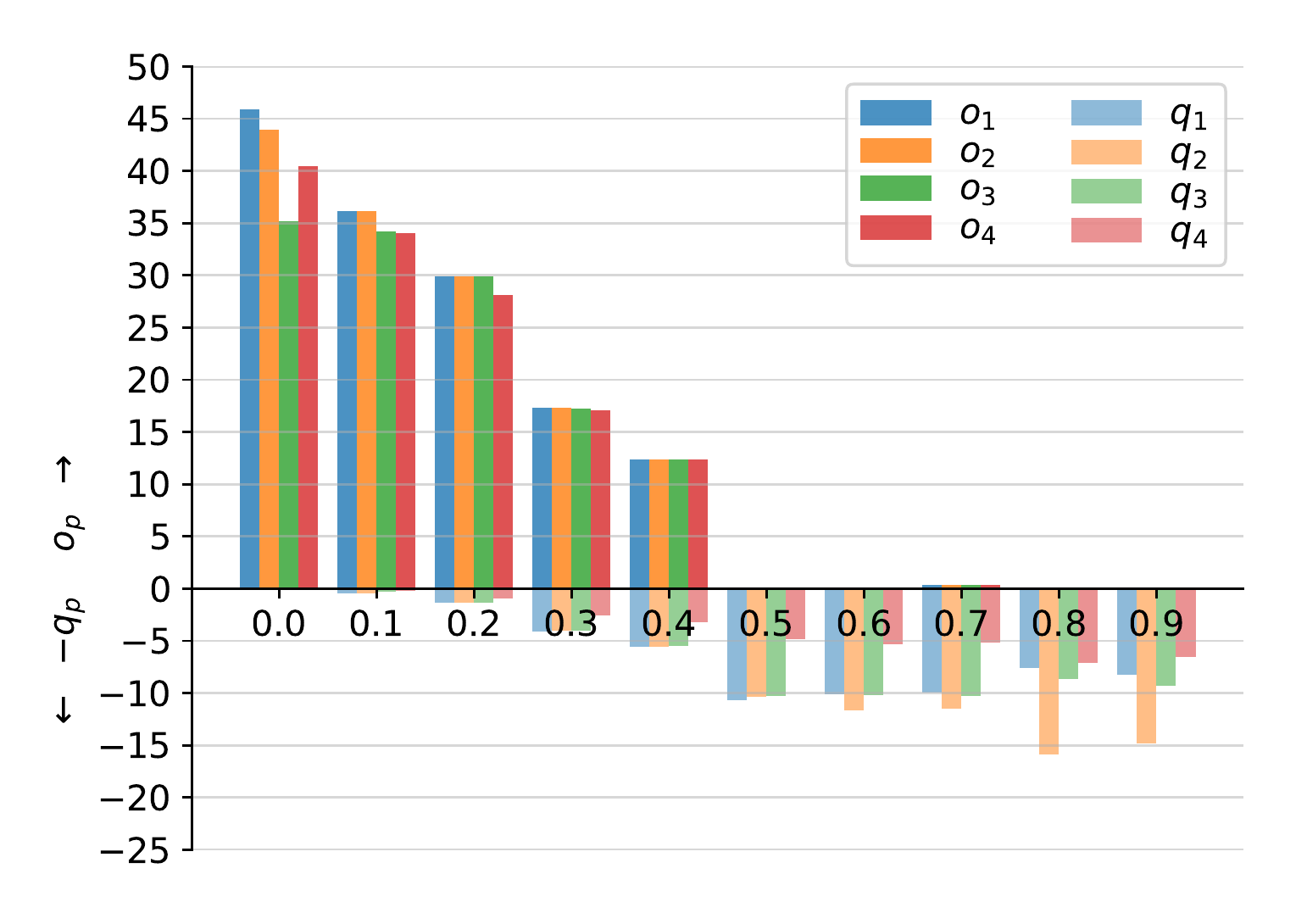}
            \caption{GHC(Gc)}
            \label{fig:objDistrAlpha4b}
        \end{subfigure}
    
        \begin{subfigure}[b]{0.5\linewidth}
     	    \centering
    	 	\includegraphics[width=\linewidth]{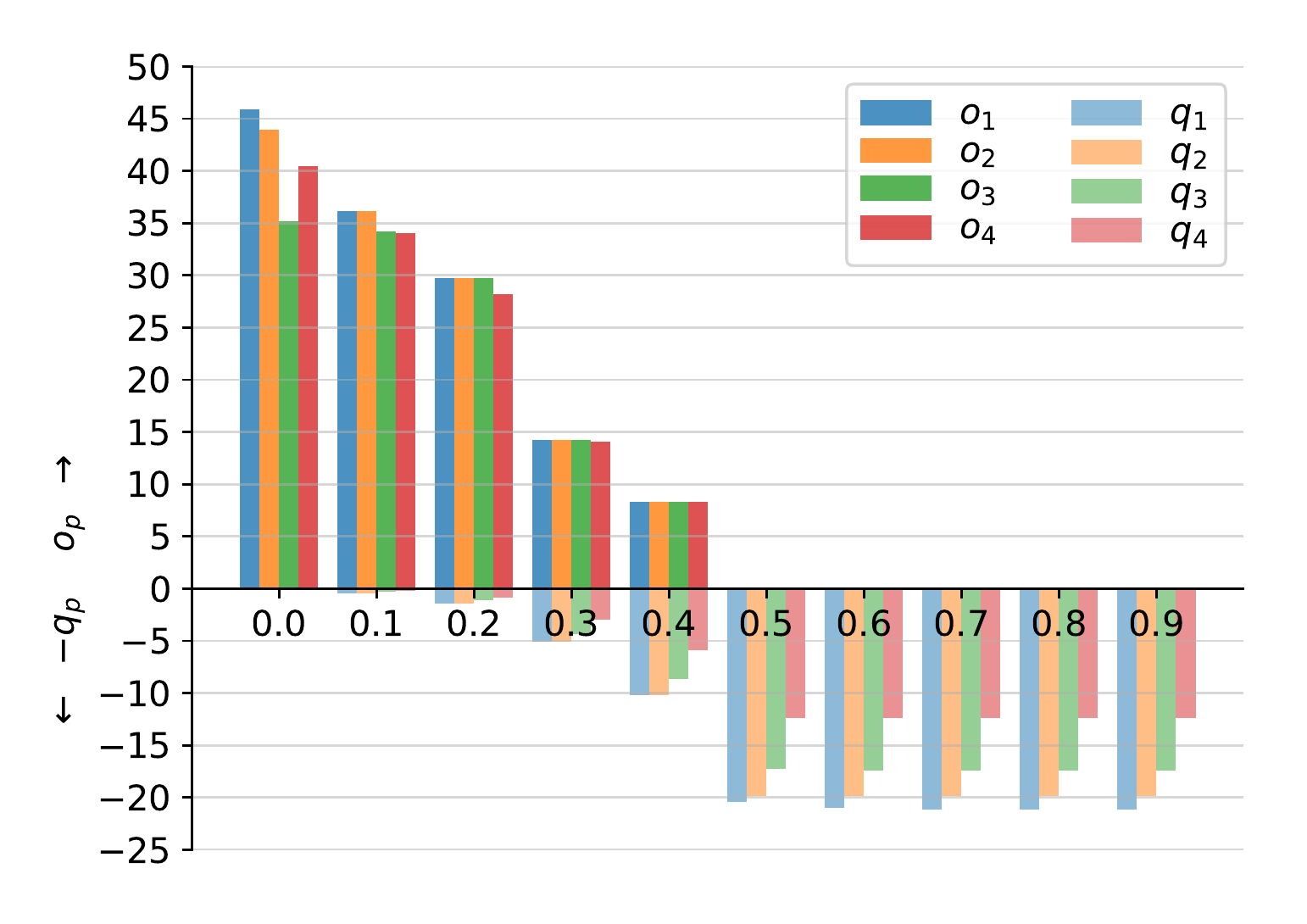}
            \caption{GHC-Inc}
            \label{fig:objDistrAlpha4c}
    	\end{subfigure}%
    	\begin{subfigure}[b]{0.5\linewidth}
     	    \centering
    	 	\includegraphics[width=\linewidth]{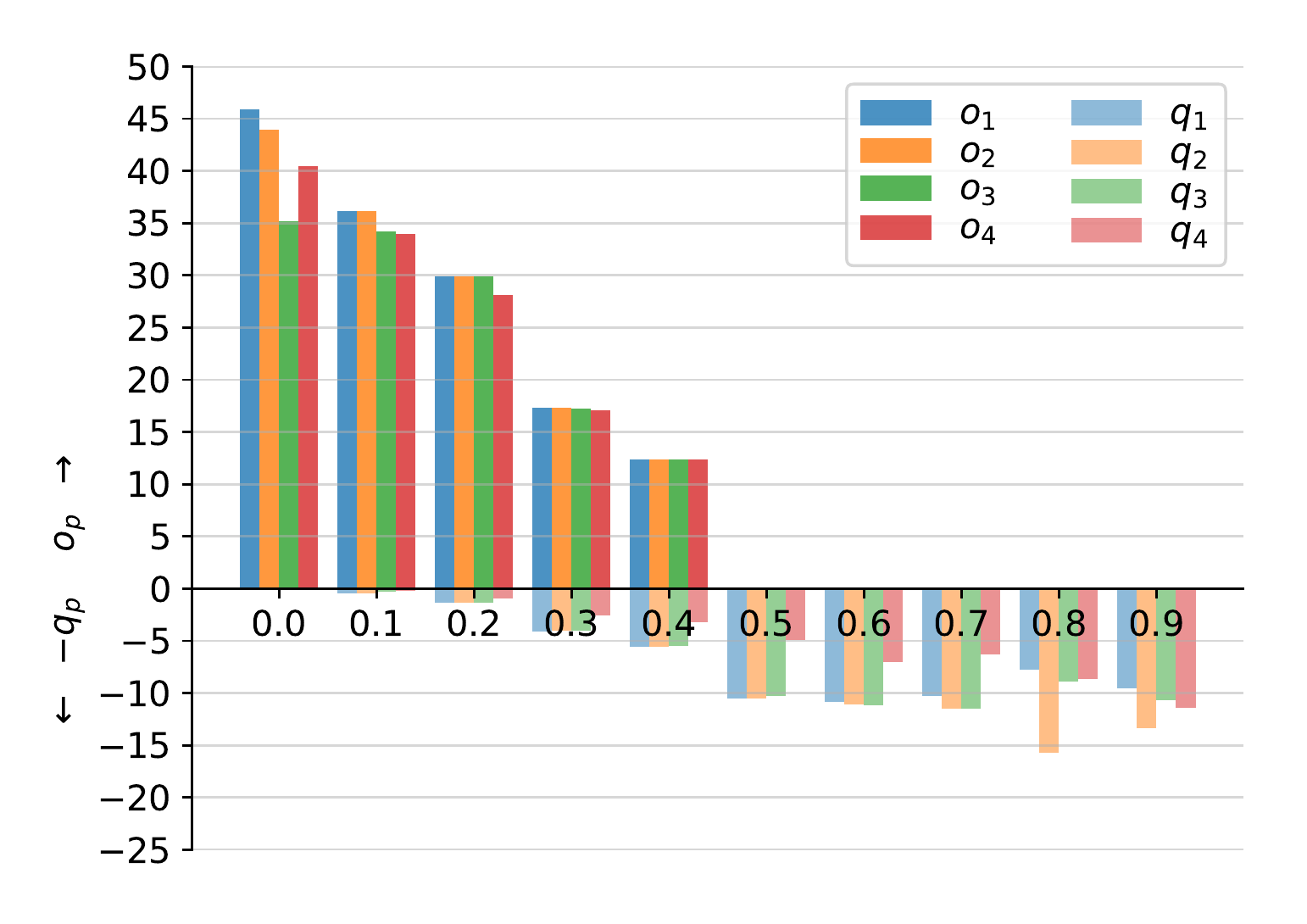}
            \caption{GHC-Tabu}
            \label{fig:objDistrAlpha4d}
    	\end{subfigure}
    
    	\caption{Distribution of the opportunity and quality objective values for different values of $\alpha$ in the difficult synthetic dataset \texttt{Gauss(1,0.3)} with four student groups. The values on the $y$-axis are multiplied by $100$ to correspond to percentages.
    	The $x$-axis represents different values of $\alpha$.
    	The positive side of the $y$-axis is the percentage of unfair recommendations w.r.t. opportunity of a student group, while the negative side shows the percentage of quality loss incurred for a student group (multiplied by -$1$). We also include the values for $\alpha=0$, which correspond to the initial values of the objectives achieved by the HSC solution, with high opportunity objective but zero quality objective.}
    	\label{fig:objDistrAlpha4}
            
    \end{figure}

\begin{figure}[tb]
        \centering
     	\begin{subfigure}[b]{0.5\linewidth}
     	    \centering
            \includegraphics[width=\linewidth]{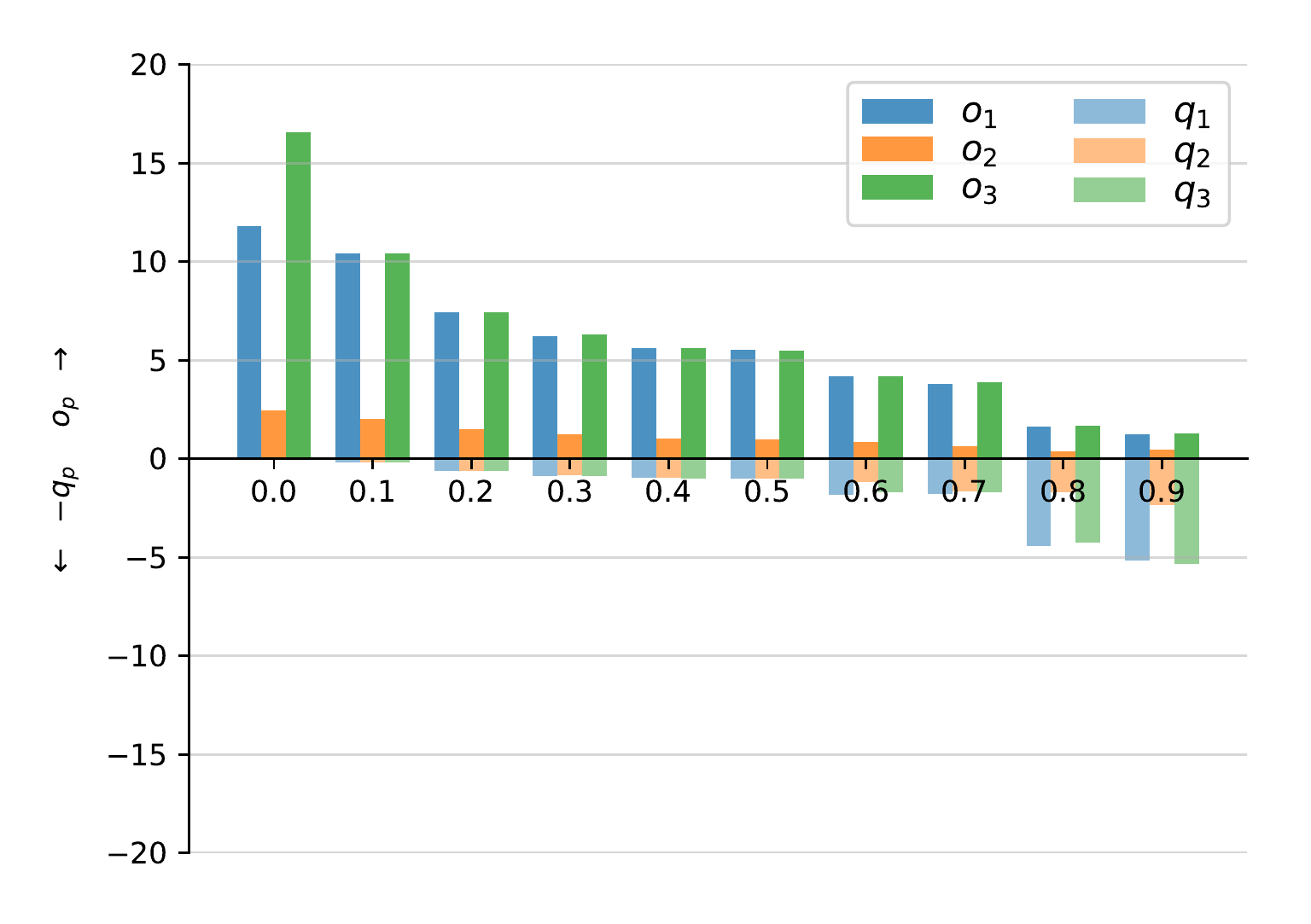}
            \caption{GHC(NoNe)}
            \label{fig:objDistrAlphaReala}
        \end{subfigure}%
        \begin{subfigure}[b]{0.5\linewidth}
     	    \centering
            \includegraphics[width=\linewidth]{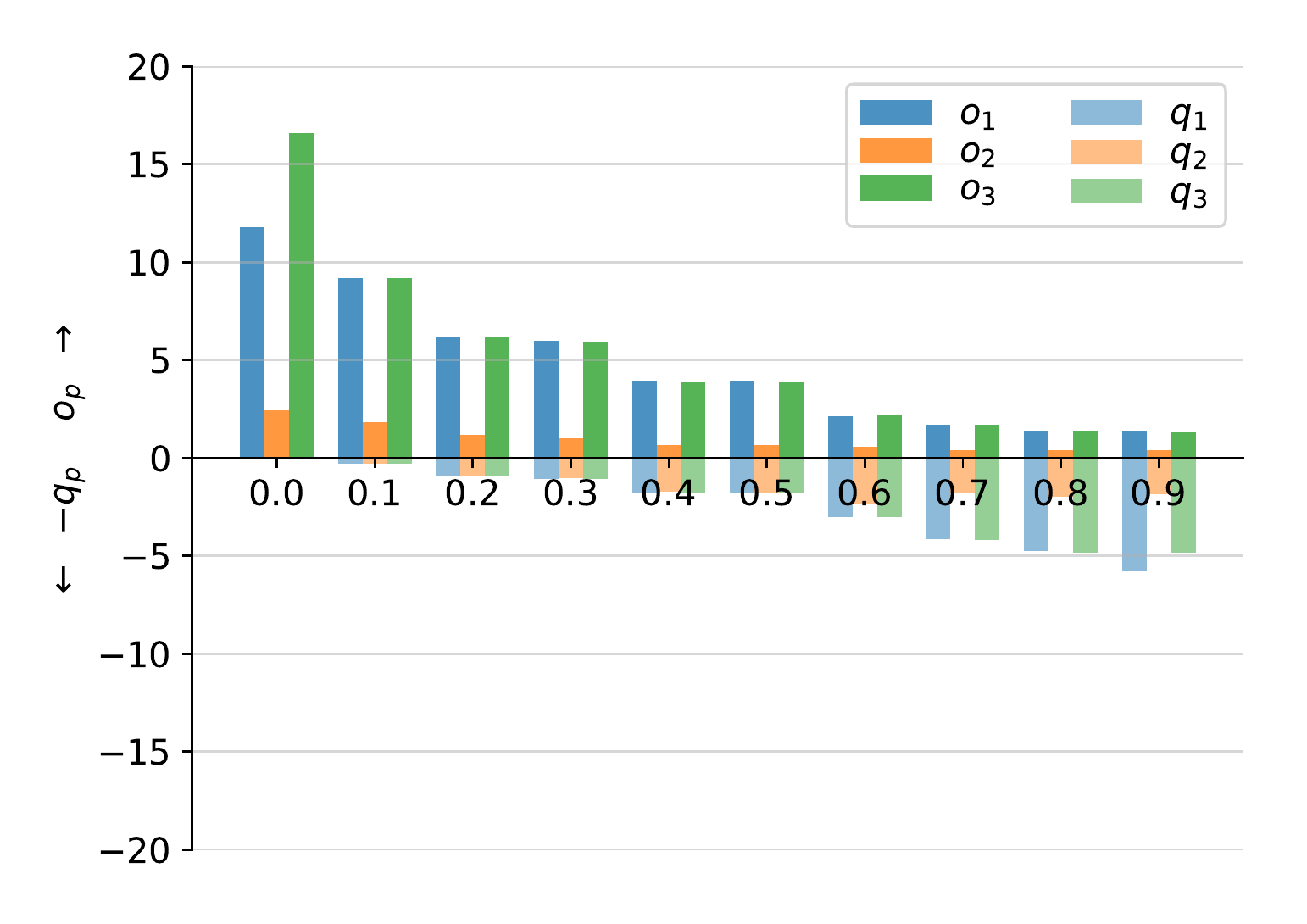}
            \caption{GHC(Gc)}
            \label{fig:objDistrAlphaRealb}
        \end{subfigure}
    
        \begin{subfigure}[b]{0.5\linewidth}
     	    \centering
    	 	\includegraphics[width=\linewidth]{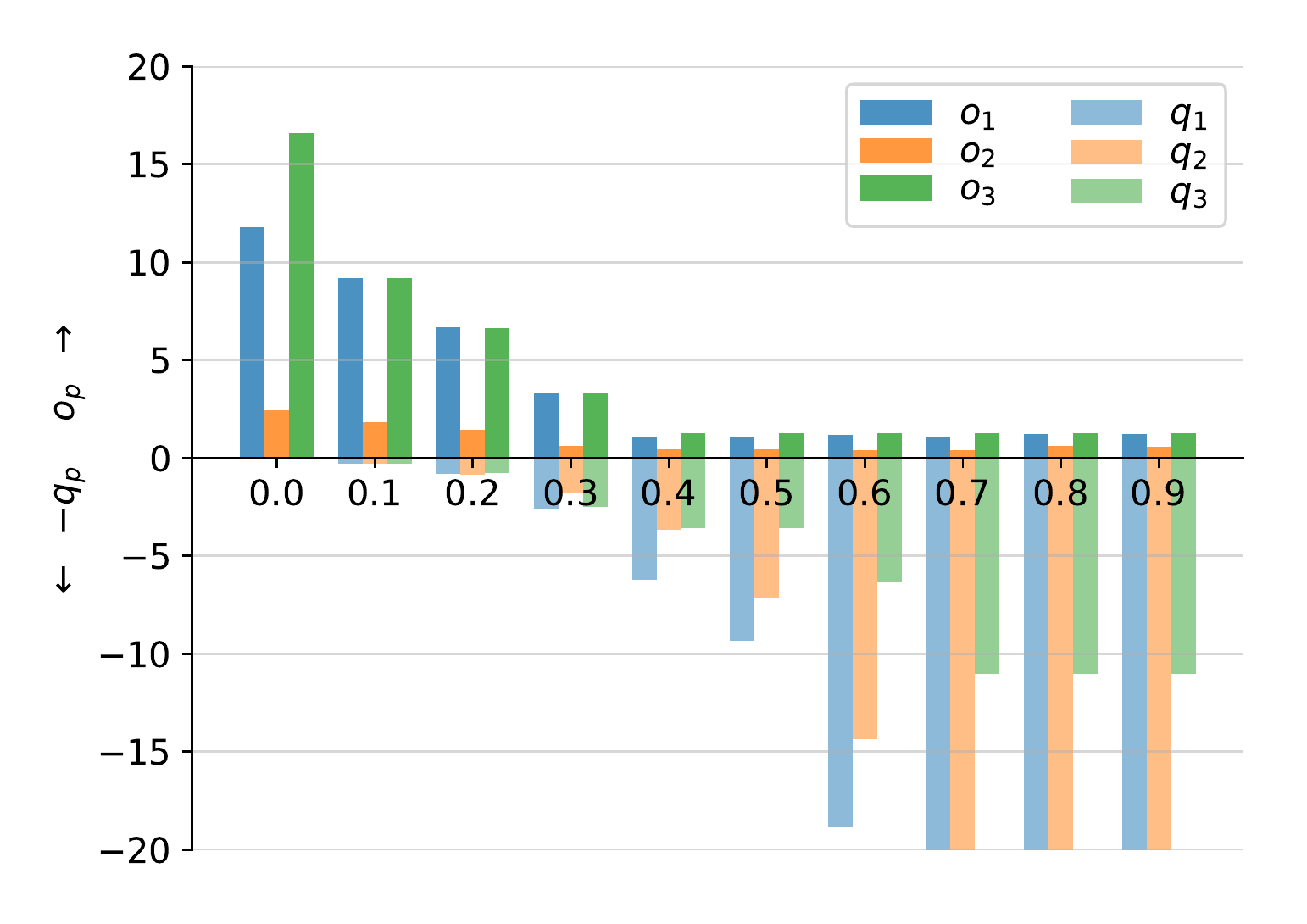}
            \caption{GHC-Inc}
            \label{fig:objDistrAlphaRealc}
    	\end{subfigure}%
    	\begin{subfigure}[b]{0.5\linewidth}
     	    \centering
    	 	\includegraphics[width=\linewidth]{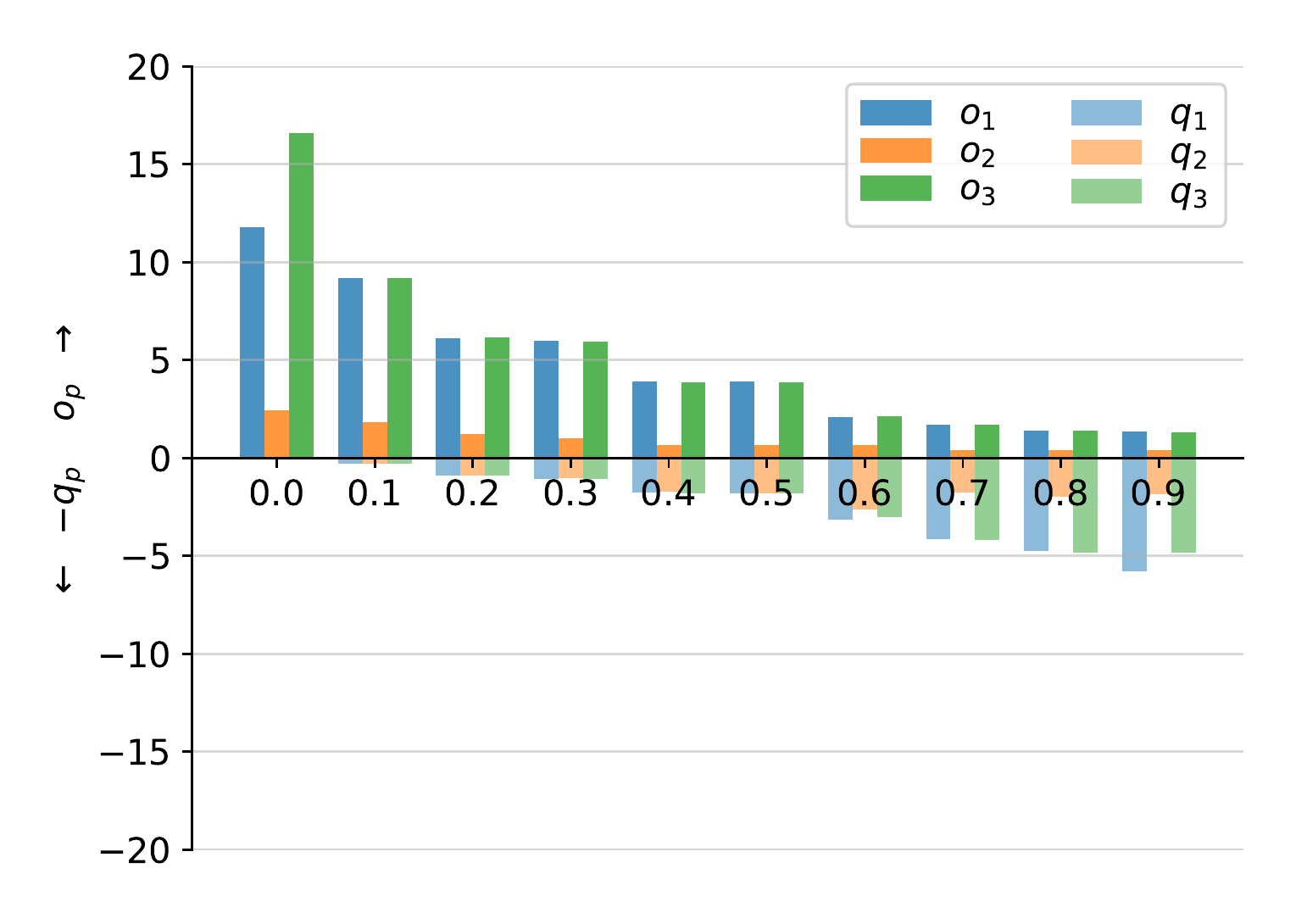}
            \caption{GHC-Tabu}
            \label{fig:objDistrAlphaReald}
    	\end{subfigure}
    
    	\caption{Distribution of the opportunity and quality objective values for different values of $\alpha$ for the \texttt{ComptSci} dataset of Fall '14. The values on the $y$-axis are multiplied by $100$ to correspond to percentages. The student groups HS, HSAP, and NAS are numbered as 1, 2, and 3 respectively.
    	The $x$-axis represents different values of $\alpha$.
    	The positive side of the $y$-axis is the percentage of unfair recommendations w.r.t. opportunity of a student group, while the negative side shows the percentage of quality loss incurred for a student group (multiplied by -$1$). We also include the values for $\alpha=0$, which correspond to the initial values of the objectives achieved by the HSC solution, with high opportunity objective but zero quality objective.}
    	\label{fig:objDistrAlphaReal}
            
    \end{figure}

The main experimental results are presented in Figures~\ref{fig:plotFQObjV} -- \ref{fig:objDistrAlphaReal}. In this section, any numbers related to the opportunity, quality, or overall objective values have been multiplied by $100$ so that the quantities correspond to percentages.  

Figure~\ref{fig:plotFQObjV} shows the scatter plots between the opportunity and quality objectives for the synthetic and \texttt{ComptSci} datasets. The $x$ axis corresponds to the highest \textbf{percentage of quality loss} that a student group may have. The $y$ axis corresponds to the highest \textbf{percentage of unfair recommendations w.r.t. opportunity} for a student group. 

Figures~\ref{fig:objDistrAlpha2}, \ref{fig:objDistrAlpha4}, and \ref{fig:objDistrAlphaReal} show how the opportunity and quality objectives are distributed across the student groups for different values of $\alpha$. Each subfigure corresponds to one of the discussed methods, GHC(NoNe), GHC(Gc), GHC-Inc, or GHC-Tabu. The positive side of the $y$-axis is the \textbf{percentage of unfair recommendations w.r.t. opportunity} of a student group, while the negative side shows the \textbf{percentage of quality loss} that a student group has. Figures~\ref{fig:objDistrAlpha2} and \ref{fig:objDistrAlpha4} correspond to the synthetic datasets \texttt{Gauss(1,0.3)} with two and four protected groups, respectively. Figure~\ref{fig:objDistrAlphaReal} refers to the \texttt{ComptSci} dataset for the Fall '14 semester. 

\subsection{Model performance}
    
    In the following paragraphs, we present the key findings regarding the performance of the different methods we evaluate w.r.t. the opportunity, quality, or overall objective values.

	\subsubsection{GHC-Tabu provides minor improvements over GHC(Gc).} 
	GHC-Tabu is an extension of GHC(Gc) which additionally performs negative moves when it reaches a local minimum, to reach a better possible solution. Based on our experimental results, GHC-Tabu performs the same or slightly better than GHC(Gc). That is the reason why we do not include the results of the GHC(Gc) algorithm in Figure~\ref{fig:plotFQObjV}. We are able to see how and when GHC-Tabu improves the solution of GHC(Gc) in Figures~\ref{fig:objDistrAlpha2}, \ref{fig:objDistrAlpha4}, \ref{fig:objDistrAlphaReal}, as they provide a more detailed view of the results. Comparing the (b) and (d) subfigures, we cannot notice any particular differences in the opportunity objective achieved by the two models. However, when $\alpha$ gets higher values, placing more weight on the opportunity objective, we see that the quality objective of the different student groups is uneven. Here is where GHC-Tabu helps. It manages to lower the quality objective by improving the objective value of the worst-performing student group. In the case of real datasets, that is harder to accomplish as we see smaller improvements of the GHC-Tabu over the GHC(NoNe). The reason for that is most likely the distribution of the recommendation scores of a student in the real datasets. In the real data, few courses will have high scores, so it is not as easy to replace them with courses that will balance the opportunity objective, if needed.
	
	\subsubsection{The proposed methods successfully improve the fairness w.r.t. the opportunity.}
	For the synthetic datasets with two protected groups, the initial percentage of unfair recommendations w.r.t. the opportunity is on average $5.2\%, 22.0\%, 44.1\%$ for the \texttt{Uni}, \texttt{Gauss(1,0.1)} and \texttt{Gauss(1,0.3)}, respectively. We manage to eliminate them with $0.1\%, 2\%, 10\%$ of quality loss, respectively. For four protected groups, we start from $10.0\%, 25.1\%, 45.0\%$ of unfair recommendations and we manage to eliminate them, while incurring only $0.5\%, 2.5\%, 10.0\%$ of quality loss, respectively. For the \texttt{ComptSci} datasets, the initial percentage of unfair recommendations w.r.t. the opportunity is $16.5\%, 14.9\%, 22.1\%$ for the Fall'14, Spring'15, and Fall'15, respectively. We manage to decrease the opportunity objective to $2\%$ or lower, with less than $10\%$ of quality degradation.
	
    \subsubsection{The difficulty of the problem increases with the number of protected groups.} 
    When there are two protected groups, for small values of $\alpha$, all methods manage to achieve $O=0$ (first row in Figure~\ref{fig:plotFQObjV-Synth}). However, when the number of protected groups increases (second row of Figure~\ref{fig:plotFQObjV-Synth} and Figure~\ref{fig:plotFQObjV-Compt}), this is not always the case. In order to achieve low values of the opportunity objective, we often need to use larger values of $\alpha$.
    
    
    \begin{figure}[tb]
 	    \centering
        \includegraphics[width=0.95\linewidth]{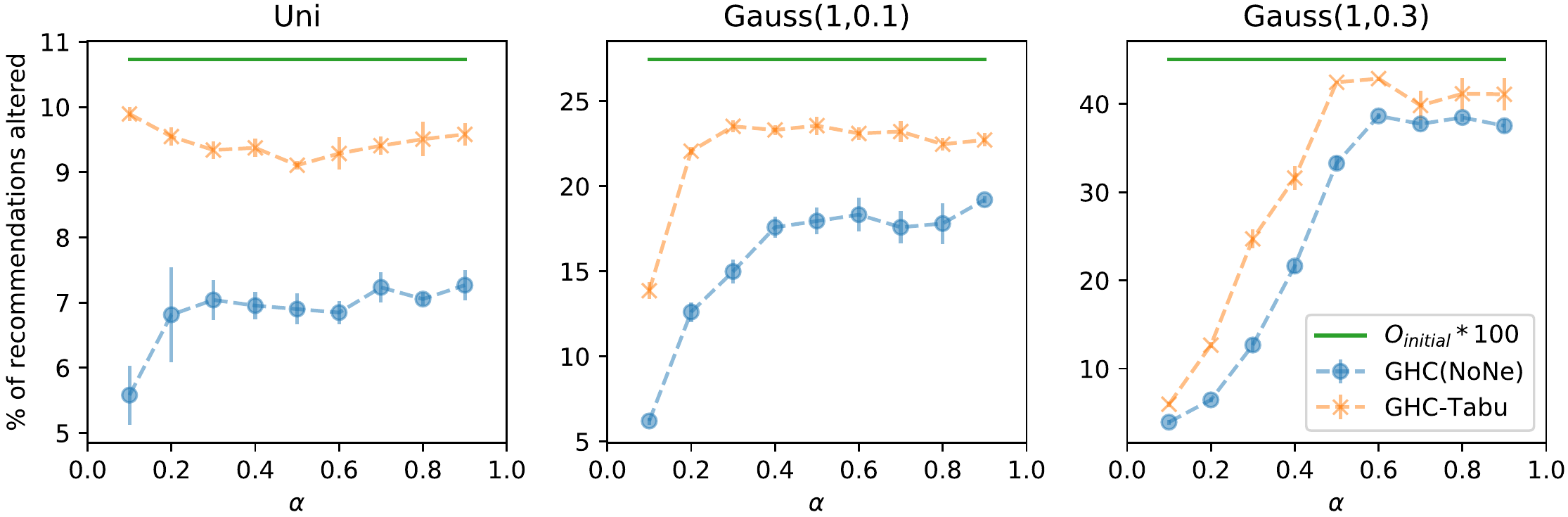}
        \caption{Percentage of recommendations affected by the algorithms in the case of four protected groups. The green horizontal line represents the \% of unfair recommendations of the most impacted group w.r.t. opportunity in the initial HSC solution.}
		\label{fig:plotSteps}
        
	 \end{figure}
    \subsubsection{Limiting the neighborhood of local search is beneficial.}  We compare GHC(NoNe) with GHC(Gc)/GHC-Tabu to evaluate the neighborhood selection presented in Sect.~\ref{sec:devalg:ghcrm:neighbor}. 
    GHC(NoNe) searches all possible moves to select the best one, while GHC-Tabu's local search is limited by the target student group and target course. 
    Their significant difference appears in the case of four protected groups (second row in Figure~\ref{fig:plotFQObjV-Synth}). For example, in the \texttt{Uni} dataset, GHC(NoNe) cannot reach values lower than 2\% for the opportunity objective, even for the highest value of $\alpha$, while GHC-Tabu achieves $O=0$. GHC-Tabu manages to better improve the solution w.r.t. the opportunity objective compared to GHC(NoNe). In particular, the easier the dataset, the worse the performance achieved by GHC(NoNe).
    By starting the local search based on the student group and the course with the highest unfairness w.r.t. opportunity, GHC-Tabu can better identify the changes that needed. On the other hand, GHC(NoNe) gets stuck easier, and reaches a local minimum without correcting enough recommendations. 
    In order to better understand this, we computed the percentage of recommendations changed by the proposed algorithms, as shown in Figure~\ref{fig:plotSteps}. 
    The easier the dataset, the less corrections GHC(NoNe) does.
    When the dataset is harder, there are more recommendations that need to change, and GHC(NoNe) performs relatively better. In any case, GHC-Tabu manages to correct more recommendations towards a more balanced outcome.
    
    Additionally, GHC-Tabu (because of GHC(Gc)) is prune on getting trapped at local minima in the case of easy datasets with unnecessary high values for $\alpha$ (last two figures in the first row of Figure~\ref{fig:plotFQObjV-Synth}). Initially, it freely and carelessly makes moves that introduce a lot of quality loss to improve the opportunity objective which is the most important term because of the high value of $\alpha$. It reaches a point where the opportunity objective is minimized, but the recommended courses have lower recommendation scores. In contrast to GHC(NoNe), GHC(Gc) searches a smaller space driven by the student groups and courses with the highest opportunity objective. As a result, it might not be able to replace the recommendations that introduce high quality loss. 
    
    For the remaining datasets with $g_s \neq 4$ in Figure~\ref{fig:plotFQObjV}, the proposed methods have similar performance, i.e., they manage to achieve the same values for the $Q$ and $O$ objectives. For example, in the \texttt{Gauss(1,0.1)} dataset for two protected groups, we see that both models achieve $O=0$, while introducing similar percentage of quality loss ($2\%$). For smaller values of $\alpha$ though, when we have higher values of the opportunity objective, we see that there is a gap between the $O$ objective achieved by GHC(NoNe) and that by GHC-Tabu ($13\%$ vs $7\%$, respectively).
    
    If we examine the results on the specific datasets presented in Figures~\ref{fig:objDistrAlpha2}, \ref{fig:objDistrAlpha4}, \ref{fig:objDistrAlphaReal}, we see that GHC(NoNe) is the worst of the four models across all cases w.r.t. the opportunity objective. Sometimes, it achieves lower percentage of quality loss, however, it has higher unbalance in course recommendations. For example, in Figure~\ref{fig:objDistrAlpha4} with the synthetic data and four protected groups, it achieves lower values for the $q_p$ compared with the rest methods, but it has higher values of $o_p$, while the other methods have $o_p=0$ for $\alpha \geq 0.5$.
	 
    \subsubsection{GHC(Inc) lowers the opportunity objective but ends up with worse quality objective for high values of alpha.} In Figure~\ref{fig:plotFQObjV}, GHC-Inc manages to reach the opportunity objective as low (or lower) as the rest of the methods. We can also see in Figures~\ref{fig:objDistrAlpha2}, \ref{fig:objDistrAlpha4}, \ref{fig:objDistrAlphaReal} that GHC-Inc performs better than GHC(NoNe), and similarly or better than the GHC-Tabu (and GHC(Gc)) in terms of the opportunity objective. Additionally, another advantage of GHC-Inc is that when it reaches $O=0$ for some value of $\alpha$, it makes no additional moves after that point and it does not degrade the quality objective any further (e.g., Figure~\ref{fig:plotFQObjV-Synth}, two protected groups, \texttt{Gauss(1,0.1)} and \texttt{Gauss(1,0.3)}, and Figures ~\ref{fig:objDistrAlpha2c}, \ref{fig:objDistrAlpha4c} with the \texttt{ComptSci} dataset). 
    
    However, if GHC-Inc achieves an opportunity objective that is close to zero, but not equal to zero, it continues to make more moves than needed, while trying to get to a slightly better local minimum. That results in an increased value of the quality objective for minor improvements in the opportunity objective. Especially in Figure~
    \ref{fig:objDistrAlphaRealc}, the values of the $q_p$ are surprisingly high. Moreover, for the datasets that GHC-Inc reaches $o_p = 0, \forall p \in [1,\cdots,g_s]$, GHC-Tabu also does the same, for the same value of $\alpha$, but with lower values of $q_p$. 
    
    This indicates that when we are interested in just reducing the opportunity objective without having a substantial quality loss, GHC-Inc is the best method to use. If we are interested in improving fairness w.r.t. opportunity as much as possible, while still maintain fairness w.r.t. recommendation quality, the best performing method is GHC-Tabu.

    \subsubsection{GHC-Tabu is the most robust model w.r.t. the input data.} 
\begin{figure}[bt]
 	    \centering
        \includegraphics[width=\linewidth]{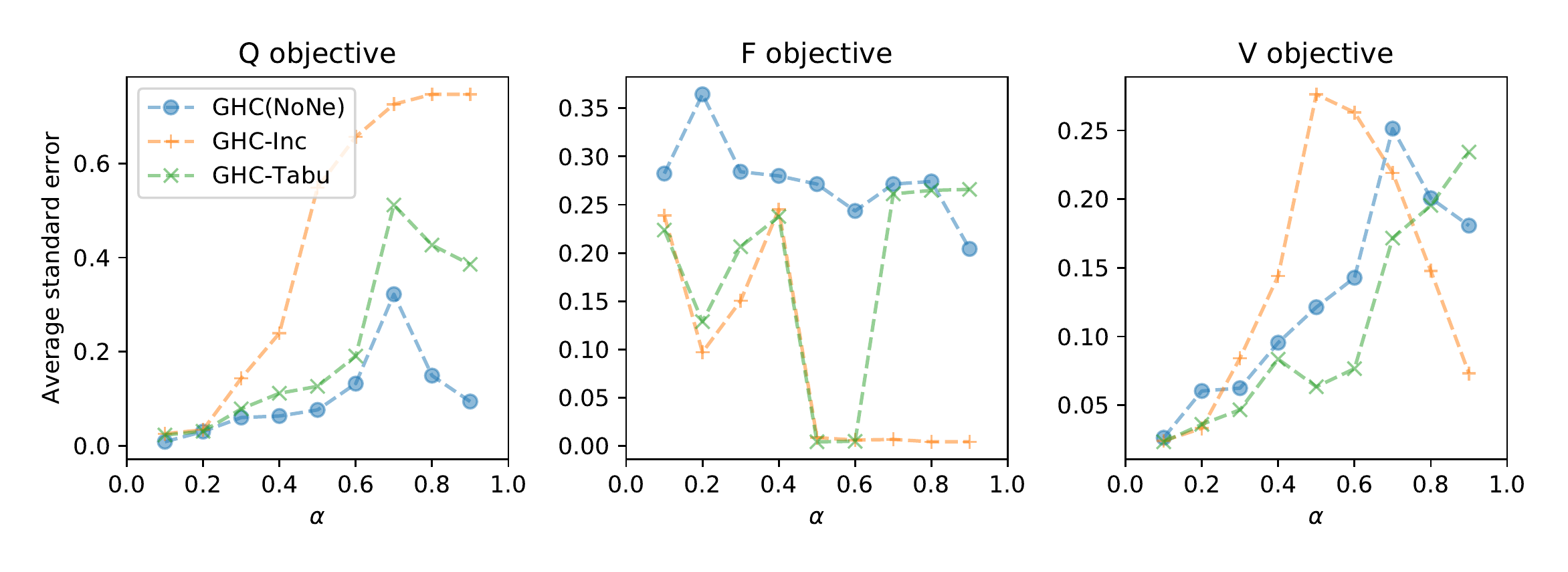}
        \caption{Average standard error of the different objective values achieved across the different families of datasets.}
		\label{fig:plotSError}
	\end{figure}

    In Figures~\ref{fig:plotFQObjV} and \ref{fig:plotSteps} for the synthetic datasets, the results are averaged over the five datasets generated with different seeds. 
    We compute the standard error (SE) of the objectives for each of those five and each value of $\alpha$. Figure~\ref{fig:plotSError} shows the averages of these SE over all the six synthetic dataset families (three difficulty levels, two and four protected student groups). These give us an indication about the robustness of the methods for datasets with similar characteristics. GHC(NoNe) is very consistent w.r.t. the $O$ objective, where it achieves $0.2$ average standard error for most of the values of $\alpha$, but less consistent regarding the quality objective. GHC-Inc has the opposite behavior, i.e., the lowest SE of the $Q$ and the highest for $O$. Overall, GHC-Tabu is more consistent as it has similar SE in the two objectives, and the lowest in the combined objective $V$.
    
	    

\subsection{The effect of parameter alpha}
\label{sec:resultsAlpha}

Figures~\ref{fig:objDistrAlpha2},~\ref{fig:objDistrAlpha4}, and~\ref{fig:objDistrAlphaReal} provide us with insights about how the value of $\alpha$ affects the distribution of the objective values across the student protected groups. In particular, the opportunity objective becomes more balanced across the student groups for higher values of $\alpha$ for all methods, as expected. Even for the real dataset in Figure~\ref{fig:objDistrAlphaReal}, GHC-Tabu achieves $o_p \leq 1.3\%$ for different values of $p$ (and standard deviation of $0.43\%$) for $\alpha = 0.9$, compared to $o_p \leq 16.5\%$ (and standard deviation of $5.87\%$) for $\alpha = 0$, which corresponds to the HSC solution. Even with $\alpha = 0.1$, GHC-Tabu manages to drop the opportunity objective values per group to $o_p \leq 9.2\%$ (and standard deviation of $3.4\%$). The higher values of $\alpha$ (together with the $L_\infty$ norm) manage to both lower the performance of the opportunity objective and balance any remaining unfair recommendations to the student groups.

For the specific dataset shown in Figure~\ref{fig:objDistrAlpha2}, because there are two protected groups of the same size, the opportunity objectives are balanced as they are complimentary, i.e., the courses that are over recommended in the one group are under recommended in the other one, and vice versa. With respect to the opportunity objective, we see some apparent difference in Figures~\ref{fig:objDistrAlpha2b} and~\ref{fig:objDistrAlpha2d}. While GHC(NoNe) and GHC-Inc are able to maintain the same $q_p$ values for higher values of $\alpha$ than needed to achieve $o_p = 0$, that is not the case for GHC(Gc) and GHC-Tabu. While GHC-Tabu has the lower $q_p$ values for $\alpha = 0.4$ with $o_p = 0$, once we set $\alpha > 0.5$, we notice a different behavior. In this case, we put more weight on the opportunity objective, so the algorithm makes some less careful moves early on that introduce high quality loss which it cannot undo afterwards. As a result, it is trapped in a local minimum, where the quality loss is not fairly distributed in the two groups. This shows that such unnecessary high values of $\alpha$ introduce unfairness with respect to the unbalanced quality objectives. That is the case in other datasets as well, but less noticeable.

\section{Conclusion}
\label{sec:summary}
	Course selection plays an important role in students' progress towards graduation, but also in the career path they will follow afterwards. In this paper, we examined group fairness in the context of course recommendation to ensure that all students are given the same opportunities when using a recommender system. We formulated a multi-objective problem that balances the fairness in opportunity and quality. We developed greedy algorithms that iteratively improve the combined objective function. The results indicate that GHC-Tabu can consistently improve fairness w.r.t. the opportunity with limited quality loss. GHC-Inc is the best method when we only assign a small weight on the opportunity objective as it gradually increases this weight in order to take more careful steps towards a fairer set of recommendations.

\backmatter

\bmhead{Acknowledgments}

This work was supported in part by NSF (1447788, 1704074, 1757916, 1834251), Army Research Office (W911NF1810344), Intel Corp, and the Digital Technology Center at the University of Minnesota. Access to research and computing facilities was provided by the Digital Technology Center and the Minnesota Supercomputing Institute.

\setlength{\bibsep}{5pt}
\bibliography{references,refmywork}



\end{document}